\documentclass[11pt,a4paper,numbers,sort&compress]{article}
\usepackage[utf8]{inputenc}
\pdfoutput=1
\usepackage{jheppub}

\usepackage{physics}
\usepackage{amsmath}
\usepackage{color}

\usepackage{slashed}
\usepackage{cancel}
\usepackage[normalem]{ulem}

\newcommand{\ul}[1]{\underline{#1}}
\newcommand{\nn}{\nonumber}
\newcommand{\fft}[2]{\frac{#1}{#2}}

\preprint{LCTP-23-07}

\title{$c$-functions in Higher-derivative Flows Across Dimensions}

\author[a]{Evan Deddo}

\author[a]{James T. Liu}

\author[a,b,c]{Leopoldo A. Pando Zayas}

\author[a]{Robert J. Saskowski}

\emailAdd{evdedd@umich.edu, jimliu@umich.edu, lpandoz@umich.edu, rsaskows@umich.edu}

\affiliation[a]{Leinweber Center for Theoretical Physics, 
University of Michigan, Ann Arbor, MI 48109, USA}

\affiliation[b]{School of Natural Sciences, Institute for Advanced Study, Princeton, NJ 08540, USA}

\affiliation[c]{The Abdus Salam International Centre for Theoretical Physics, 34014 Trieste, Italy}

\abstract{In the context of gravitational theories describing renormalization group flows across dimensions via AdS/CFT,  we study the role of higher-derivative corrections to Einstein gravity. We use the Null Energy Condition to derive monotonicity properties of candidate holographic central charges formed by combinations of metric functions. We also implement an entropic approach to the characterization of the four-derivative flows using the Jacobson-Myers functional  and demonstrate, under reasonable conditions, monotonicity of certain terms in the entanglement entropy  via the appropriate generalization of the Ryu-Takayanagi prescription. In particular, we show that any flow from a higher dimensional theory to a holographic CFT$_2$ satisfies a type of monotonicity. We also uncover direct relations between NEC-motivated and entropic central charges. }

\date{\today}

\begin{document}

\maketitle
\section{Introduction  and summary}
One central organizing principle in the space of quantum field theories (QFTs) is the Renormalization Group (RG) flow. RG flow is often understood as a family of successive quantum field theories starting at some high-energy (UV) conformal field theory (CFT) and flowing to some low-energy (IR) CFT. As the flow progresses, the effective number of degrees of freedom decreases due to the process of coarse-graining. This reduction can be accurately quantified by ``counting functions,'' which are monotonic along the RG flow and thus render the flows irreversible. Of particular interest are functions that connect quantities in the CFTs, such as $A$-type central charges in even dimensions and sphere free energies in odd dimensions. Both of these quantities will be referred to as central charges in the following discussion. There are well-established theorems regarding such flows, including proofs of the 2d $c$-theorem by Zamolodchikov \cite{Zamolodchikov:1986gt}, the 3d $F$-theorem by Casini and Huerta \cite{Jafferis:2010un,Klebanov:2011gs,Jafferis:2011zi,Casini:2012ei}, and the 4d $a$-theorem by Komargodski and Schwimmer \cite{Cardy:1988cwa,Komargodski:2011vj}; an alternative approach that has been used to great effect involves entanglement entropy and has been quite useful for proving results in $d=2,3,4$ \cite{Casini:2006es,Casini:2017roe,Casini:2017vbe}. There also exist partial results in 5d \cite{Jafferis:2012iv,Chang:2017cdx,Fluder:2020pym} and 6d \cite{Elvang:2012st,Cordova:2015fha,Heckman:2015axa}.

The AdS/CFT correspondence geometrizes many aspects of QFTs and has proven a particularly useful framework to study the properties of RG flows. Considerable progress on constructing $c$-functions has been made from the holographic perspective:  Various holographic $c$-theorems have been established in this context by making use of the Null Energy Condition (NEC) \cite{Girardello:1998pd,Freedman:1999gp,Myers:2010tj,Myers:2010xs}, as well as using the entanglement entropy perspective to analyze holographic RG flows \cite{Casini:2011kv,Myers:2012ed}. Holographic methods, for example, permit the construction of certain monotonic $c$-functions in any dimension and at strong coupling, something way beyond the reach of  field-theoretic approaches.

Naturally, much work has been done on extending holographic $c$-theorems to include higher-derivative corrections \cite{Anber:2008js,Myers:2010tj,Myers:2010xs,Myers:2012ed,Sinha:2010ai,Gullu:2010pc,Gullu:2010st,Sinha:2010pm,Oliva:2010eb,Myers:2010ru,Oliva:2010zd,Liu:2010xc,Liu:2011iia,Alkac:2018whk,Ghodsi:2019xrx,Anastasiou:2021jcv,Ghodsi:2021xrb,Alkac:2022zda}.  Such extensions allow one to distinguish various central charges \cite{Nojiri:1999mh, Blau:1999vz}. For example, in 4D, we have that $a=c$ at the two-derivative level in gravity or in the large-$N$ limit in field theory. It is well known, however, that $a$ alone has a monotonic flow from the UV to the IR \cite{Komargodski:2011vj}, while $c$ does not. As such, adding higher derivatives allows one to distinguish between the central charges that have monotonic flows and the ones that do not.  Such higher derivatives correspond to sub-leading in $N$ corrections to the central charges. 

In this work we explore the notion of counting functions in RG flows across dimensions, meaning the compactification of a $D$-dimensional CFT, which is the UV fixed point, on a $(D-d)$-dimensional compact space, such that the IR fixed point is a $d$-dimensional CFT.  RG flows across dimensions are particularly amenable to holographic methods; there are many examples of supergravity solutions holographically dual to RG flows interpolating between CFTs of different dimensions \cite{Maldacena:2000mw,Acharya:2000mu,Gauntlett:2000ng,Gauntlett:2001jj,Gauntlett:2001qs,Benini:2013cda,Benini:2015bwz,Bobev:2017uzs}. Some candidate $c$-functions for such flows were studied in \cite{Macpherson:2014eza,Bea:2015fja,Legramandi:2021aqv}, and more recently an explicit $c$-function was constructed in \cite{GonzalezLezcano:2022mcd}. The holographic entanglement entropy picture for such flows was further analyzed in \cite{Deddo:2022wxj}.  In this manuscript, we explore the role of higher-derivative corrections in holographic flows across dimensions. As a natural starting point, we generalize some of the results of Myers-Sinha \cite{Myers:2010xs,Myers:2010tj}, who considered the effect of higher-derivative terms in holographic RG flows, to flows across dimensions. 

\subsection{The holographic setup}

Our starting point is Einstein gravity with a negative cosmological constant.  From an effective field theory point of view, one would expect this to be corrected by a set of higher derivative operators.  The first such terms arise at the four-derivative level and involve a combination of $\hat R_{MNPQ}\hat R^{MNPQ}$, $\hat R_{MN}\hat R^{MN}$, and $\hat R^2$.  However, since the Ricci terms can be shifted by a field redefinition, we may choose the Gauss-Bonnet combination
\begin{equation}
    \chi_4=\hat R_{MNPQ}\hat R^{MNPQ}-4\hat R_{MN}\hat R^{MN}+\hat R^2.
\end{equation}
As a result, we focus on the bulk $(D+1)$-dimensional Lagrangian
\begin{equation}
    e^{-1}\mathcal{L}=\frac{1}{2\kappa^2}\qty[\hat R+\frac{D(D-1)}{L^2}+\alpha\chi_4],
\label{eq:gravLag}
\end{equation}
where $\alpha$ parametrizes the correction.  This choice of the Gauss-Bonnet combination is convenient since in this case, the corrected Einstein equation remains second order in derivatives.  This system admits a maximally symmetric AdS$_{D+1}$ vacuum with an AdS radius $L_{\mathrm{UV}}^2=L^2-\alpha(D-2)(D-3)$, to linear order in $\alpha$.

We are interested in flows from AdS$_{D+1}$ in the UV to AdS$_{d+1}\times M_{D-d}$ in the IR.  Such flows can be induced by coupling the gravitational Lagrangian, (\ref{eq:gravLag}), to a suitable matter sector, \textit{i.e.}, $\mathcal L\to\mathcal L+\mathcal L_{\mathrm{matter}}$.  To parametrize the flow, we split off the would-be internal space $M_{D-d}$ and assume a spacetime metric of the form
\begin{equation}
        \dd s^2=e^{2f(z)}(\eta_{\mu\nu}\dd x^\mu \dd x^\nu+ \dd z^2)+ e^{2g(z)}g_{ij}(y)\dd y^i\dd y^j.
\end{equation}
The flow is along the bulk radial coordinate, $z$, and we take the asymptotics to be such that $e^{2f}\sim e^{2g}\sim 1/z^2$ in the UV ($z\to0$) while $e^{2f}\sim 1/z^2$ with $e^{2g}\sim\mbox{const.}$ in the IR ($z\to\infty$). Note that this metric implicitly assumes flat slicings of $\mathrm{AdS}_{d+1}$, although some authors have considered curved slicings \cite{Ghosh:2017big,Ghosh:2018qtg,Kiritsis:2022oww}.

Given a bulk metric parametrized by the two functions $f(z)$ and $g(z)$, we then explicitly construct a function $c(f,g; z)$ such that $\dd c/\dd z\le 0$ upon imposing the NEC, $T_{MN}\xi^M\xi^N\ge 0$, on the matter sector where $\xi$ is a future-directed null vector. This is the desired monotonicity property. This $c$-function directly generalizes the two-derivative case \cite{GonzalezLezcano:2022mcd} to which it reduces when the Gauss-Bonnet coupling $\alpha$ is sent to zero, as well as generalizing the four-derivative case of flows within the same dimension \cite{Myers:2010tj,Myers:2010xs} to which it reduces in the limit that there are no compact internal dimensions. This $c$-function is not unique, but instead has two free parameters characterizing it; despite this mild ambiguity, the IR limit of this central charge is unambiguously the $A$-type central charge as expected.  In other words, $\lim_{z\to\infty}c(z)= a_\text{IR}$, where $a_{\text{IR}}$ is the four-derivative $A$-type central charge.

As in the two-derivative case \cite{GonzalezLezcano:2022mcd}, this $c$-function diverges in the UV.  However, we show that the divergence of the $c$-function encodes the UV central charge. As we approach the UV, the compact extra dimensions unfurl and our massive KK towers become increasingly light and begin to enter the spectrum, meaning that the number of lower-dimensional degrees of freedom appears to become infinite. Dimensional analysis alone tells us that the central charge must diverge as a pole of order the number of compact dimensions; however, we go further and show that the coefficient of this pole encodes the value of the UV central charge, \textit{i.e.},
\begin{equation}
    c(z)\overset{z\to 0}{\sim}\frac{a_\text{UV}}{z^{D-d}},
\end{equation}
where $a_\text{UV}$ is the (four-derivative) $A$-type central charge in the UV. This is not entirely automatic; it requires an additional constraint on the remaining free parameters of the $c$-function. However, we may always choose the parameters so that this is the case.

We also construct $c$-functions from the entanglement entropy. In particular, we consider entangling regions of this CFT which completely wrap the internal space. The entanglement entropy has been shown \cite{Faulkner:2013ana,Hung:2011xb,deBoer:2011wk,Bhattacharyya:2013jma} to then be given by finding the extremal surface which minimizes the Jacobson-Myers functional \cite{Jacobson:1993xs}
    \begin{equation}
    S_\text{JM}=\frac{1}{4G_N}\int_\Sigma \dd[d] x \sqrt{h}\left(1+2\alpha {\cal R}\right)+\frac{1}{2G_N}\int_{\partial\Sigma}\dd[d-1]{x}\sqrt{\tilde h}(2\alpha{\cal K}),
\end{equation}
where $\Sigma$ is the extremal surface with boundary $\partial\Sigma$, $h$ is the determinant of the induced metric on $\Sigma$, $\tilde h$ is the induced metric (of the induced metric $h$) on $\partial\Sigma$, ${\cal R}$ is the scalar curvature of $\Sigma$, and $\cal K$ is the trace of the extrinsic curvature of the boundary $\partial\Sigma$.
    For the case of flows from AdS$_{D+1}$ to AdS$_3$, which may be equivalently viewed as flows from CFT$_D$ to CFT$_2$, we explicitly obtain a monotonic $c$-function from the entanglement entropy as
    \begin{equation}
        c_\text{EE}=R\,\partial_R \,\, S_\text{JM},\label{eq:monoC}
    \end{equation}
    where $R$ is the radius of the entangling region. Given a minimal surface whose profile is $r(z)$, $S_\text{JM}$ admits a first integral that can be solved for $r'(z)$. This then allows us to explicitly evaluate \eqref{eq:monoC} and subsequently verify its monotonicity as a consequence of the NEC. Moreover, it turns out to be the case that this $c$-function which one obtains from the holographic entanglement entropy is indeed related to the local $c$-function obtained directly from the NEC; we show that the monotonicity of one directly implies the monotonicity of the other. Such precise connection of two {\it a priori} differently defined $c$-functions opens the possibility of better understanding the connection between strong subadditivity of the entanglement entropy and the NEC as a condition on the holographic gravity backgrounds. 

    The rest of this paper is organized as follows: In Section \ref{sec:LH}, we explicitly construct a local $c$-function for the case of Gauss-Bonnet corrected gravity and demonstrate that it flows monotonically from the UV to the IR as a consequence of the null energy condition (NEC). In Section \ref{sec:limits}, we show that the IR limit of this $c$-function is the $A$-type central charge and, although the $c$-function diverges in the UV, the coefficient of this divergence encodes the UV central charge. In Section \ref{sec:EE}, we discuss the $c$-function obtained from holographic entanglement entropy and show that it is monotonic, at least when there is no curvature of the internal space, and show that this quantity is related to the NEC-motivated central charge constructed in Section \ref{sec:LH}. A summary and conclusions are given in Section \ref{sec:end}. We relegate some of the more technical details to Appendix \ref{App:Geom}.

\section{Higher-derivative gravity and  NEC}\label{sec:LH}

We are interested in RG flows from CFT$_D$ to CFT$_d$ triggered by compactification on a $(D-d)$-dimensional manifold, $M_{D-d}$. Holographically, this corresponds to a geometric flow from AdS$_{D+1}$ to AdS$_{d+1}\times M_{D-d}$. The holographic radial coordinate $z$ then naturally functions as the scale for RG flow. We may explicitly realize this setup by choosing a metric
\begin{equation}
\label{Eq:Flowfg}
    \dd s^2 = e^{2f(z)}(\eta_{\mu\nu}\dd x^\mu \dd x^\nu+ \dd z^2)+ e^{2g(z)}g_{ij}(y)\dd y^i\dd y^j,
\end{equation}
such that in the UV region $z\to 0$ the metric is asymptotically AdS$_{D+1}$ and in the IR region $z\to\infty$ the metric asymptotes to AdS$_{d+1}\times M_{D-d}$. To be rigorous, the metric \eqref{Eq:Flowfg} is not the most general metric describing holographic
RG flows across dimensions; for example, there are known holographic RG flows where the
internal space $M_{D-d}$ depends on the holographic radial coordinate $z$ in a non-separable way \cite{Anderson:2011cz,Fluder:2017nww,Bobev:2020jlb}. We restrict our attention to the separable case \eqref{Eq:Flowfg} for simplicity; we leave it as an exercise for future research to extend our analysis of holographic $c$-functions in separable flows to more general non-separable flows.

Furthermore, unless otherwise specified, we will assume that the metric $g_{ij}$ of $M_{D-d}$ is maximally symmetric with Ricci curvature
\begin{equation}
    \tilde R_{ij}=\kappa\frac{D-d-1}{\ell^2}g_{ij},
\end{equation}
where $\kappa=-1$, $0$, or $1$ for negative, flat, or positive curvature, respectively. This is not the most general choice of metric on the internal space but we make this choice for simplicity; we will generalize this to arbitrary Einstein internal manifolds in Section \ref{sec:generalization}.

As discussed above, we start with a two-derivative theory in the gravitational sector, namely the Einstein-Hilbert Lagrangian with a negative cosmological constant.  At the four-derivative level, we add a Gauss-Bonnet coupling
\begin{equation}
    \chi_4=\hat R_{MNPQ}\hat R^{MNPQ}-4\hat R_{MN}\hat R^{MN}+\hat R^2,
\label{eq:chi4}
\end{equation}
so we end up considering the gravitational Lagrangian
\begin{equation}
    e^{-1}\mathcal{L}=\frac{1}{2\kappa^2}\qty[\hat R+\frac{D(D-1)}{L^2}+\alpha\chi_4],
\label{eq:Lgrav}
\end{equation}
coupled to a matter sector satisfying the null energy condition.

While the NEC is a condition on the matter, namely $T_{MN}\xi^M\xi^N\ge0,$ with $\xi$ a future-directed null vector, the Einstein equation allows this to be recast as a condition on the four-derivative corrected geometry, namely
\begin{equation}
    \qty[\hat R_{MN}+\alpha\qty(\hat R_{MPQR}\hat R_{N}^{\ \ PQR}-2\hat R^{PQ}\hat R_{MPNQ}-2\hat R_{MP}\hat R_N^{\ \ P}+\hat R \hat R_{MN})]\xi^M\xi^N\ge0.
\label{eq:GeneralNEC}
\end{equation}
The main result of this section is to show that the NEC \eqref{eq:GeneralNEC} implies the existence of a monotonic $c$-function from the UV to the IR in the background \eqref{Eq:Flowfg}.


\subsection{Domain wall flows}
Before discussing flows across dimensions, let us first review the case of flows within the same dimension \cite{Myers:2010tj,Myers:2010xs}, \textit{i.e.}, for which we have a metric of the form
\begin{equation}
    \dd{s}^2=e^{2f(z)}\qty(\eta_{\mu\nu}\dd x^\mu\dd x^\nu+\dd z^2).
\end{equation}
Pure AdS corresponds to the solution $f(z)=\log(L/z)$, with $L$ being the AdS radius. Then, in these coordinates, $z=0$ corresponds to the UV and $z=\infty$ corresponds to the IR. Thus, we have a gravity solution that is a domain wall interpolating between two AdS$_{D+1}$ regions; the corresponding field theory interpretation is that of an RG flow \cite{Girardello:1998pd,Freedman:1999gp}. One can calculate the curvature tensor components
\begin{align}
    \hat R_{\mu\nu\rho\sigma}&=-e^{-2f}(f')^2\qty(\eta_{\mu\rho}\eta_{\nu\sigma}-\eta_{\mu\sigma}\eta_{\nu\rho}),&\hat R_{\mu z\nu z}&=-e^{2f}f''\eta_{\mu\nu},\nonumber\\
    \hat R_{\mu\nu}&=-\qty[f''+(D-1)(f')^2]\eta_{\mu\nu},&\hat R_{zz}&=-Df''.
\end{align}
Choosing a null vector $\xi=\partial_t\pm\partial_z$, the NEC with Gauss-Bonnet corrections is then simply expressed as \cite{Myers:2010tj,Myers:2010xs}
\begin{equation}
    (D-1)\qty(e^{-f})''\qty(1-2\alpha(D-2)(D-3) e^{-2f} (f')^2)\ge0.
\label{eq:dwNEC}
\end{equation}
Note that this will be the only non-trivial NEC due to the planar symmetry of the domain wall.

We now consider flows to the IR.  In the IR, the $A$-type central charge may be computed via the methods of \cite{Henningson:1998gx,Henningson:1998ey} to be \cite{Myers:2010jv}
\begin{equation}
    a_\text{IR}=\frac{L_\text{IR}^{D-1}}{G_N}\qty(1-2(D-1)(D-2)\frac{\alpha}{L_\text{IR}^2}),
\end{equation}
where $G_N$ is the $(D+1)$-dimensional Newton's constant.  In order to obtain a $c$-function, note that in the IR, we expect that $e^f\sim L_\text{IR}/z$, so that $(e^{-f})'\sim 1/L_\text{IR}$.  Replacing $L_{\mathrm{IR}}$ by an effective AdS radius
\begin{equation}
    L_\text{eff}(z)=\frac{1}{(e^{-f})'},
\end{equation}
that interpolates between $L_\text{UV}$ and $L_\text{IR}$ then leads to a natural ansatz for an unnormalized $c$-function
\begin{equation}
    c(z)=\frac{1}{G_N\qty(\qty(e^{-f})')^{D-1}}\qty(1-2\alpha(D-1)(D-2)((e^{-f})')^2).
\label{eq:dwcfnct}
\end{equation}
Taking a derivative with respect to $z$, one gets that
\begin{equation}
    c'(z)=-\frac{(D-1)\qty(e^{-f})''}{G_N\qty(\qty(e^{-f})')^{D}}\qty(1-2\alpha(D-2)(D-3)((e^{-f})')^2)\le 0,
\end{equation}
where the final step makes use of the null energy condition, (\ref{eq:dwNEC}). So, there is a monotonically non-increasing flow of $c(z)$ from the UV to the IR. Moreover, one can check that this function $c(z)$ interpolates between the UV and IR central charges, in the sense that
\begin{equation}
    c(z=\infty)=a_\text{IR},\qquad c(z=0)=a_\text{UV}.
\end{equation}
Here $a_\text{IR}$ and $a_\text{UV}$ are the $A$-type central charges in the IR and UV, respectively, where \begin{equation}
    a_\text{UV}=\frac{L_\text{UV}^{D-1}}{G_N}\qty(1-2(D-1)(D-2)\frac{\alpha}{L_\text{UV}^2}).
\end{equation}
This expression agrees with \cite{Myers:2010jv}.
\subsection{Two-derivative flows across dimensions}

We now turn to the case at hand, which is flows across dimensions.  Before considering the full case, we start by reviewing the two-derivative case of flows across dimensions \cite{GonzalezLezcano:2022mcd}, {\it i.e.}, without higher-derivative corrections. For such flows, we use the full metric ansatz \eqref{Eq:Flowfg}, with corresponding Ricci tensor components
\begin{align}
    \hat R_\nu^\mu&=-e^{-2f}\qty[f''+f'\qty((d-1)f'+(D-d)g')]\delta_\nu^\mu,\nonumber\\
    \hat R^i_j&=e^{-2g}\tilde R^i_j-e^{-2f}\qty[g''+g'\qty((d-1)f'+(D-d)g')]\delta^i_j,\nonumber\\
    \hat R^z_z&=-e^{-2f}\qty[df''+(D-d)\qty(g''+g'(g'-f'))].
\end{align}
Note that because we are assuming the AdS$_{d+1}$ in the IR to have flat slicings, the corresponding Ricci tensor $R_{\mu\nu}$ will vanish.

At the two-derivative level, the null energy condition is equivalent to $R_{MN}\xi^M\xi^N\ge0$.  Since the $D$-dimensional isometry is broken by the flow, we end up with two independent inequalities, which correspond to choosing null vectors along $t$-$z$ and $t$-$y$.  These conditions are, respectively,
\begin{subequations}
    \begin{align}
    \text{NEC1: }&-(d-1)\qty(f''-(f')^2)-(D-d)\qty(g''+g'(g'-2f'))&\ge0,\\
    \text{NEC2: }&(f'-g')'+(f'-g')\qty((d-1)f'+(D-d)g')+\kappa\frac{D-d-1}{\ell^2}e^{2f-2g}&\ge0.
\end{align}\label{eq:twoDerivNEC}%
\end{subequations}
NEC1 may be suggestively rewritten as
\begin{equation}
    \qty(e^{-\tilde f})''\ge \frac{(D-1)(D-d)}{(d-1)^2}e^{-\tilde f}(g')^2\ge0,
\label{eq:NEC1rew}
\end{equation}
where $\tilde f$ is an effective warp factor
\begin{equation}
    \tilde f(z)\equiv f(z)+\frac{D-d}{d-1}g(z).
\end{equation}
Likewise, NEC2 can be rearranged into the form
\begin{equation}
    \qty(e^{(d-1)f+(D-d)g}(f'-g'))'\ge -\kappa\frac{D-d-1}{\ell^2}e^{(d+1)f+(D-d-2)g}.
\label{eq:NEC2form}
\end{equation}
Note that the sign of the right-hand side term depends on the sign of the internal curvature, $\kappa$.  For $\kappa=-1$ or $\kappa=0$ the expression on the left-hand side is non-negative.  But for $\kappa=1$ the sign of this term is unconstrained.

As in the domain wall flow, we seek a $c$-function that flows to $a_{\mathrm{IR}}$ in the IR.  Before constructing such a function, we first recall the asymptotics of the flow.  Flowing from AdS$_{D+1}$ in the UV to AdS$_{d+1}$ in the IR, one expects
\begin{align}
    \text{UV }(z=0)&:\ \ (e^{-f})'=(e^{-g})'=\frac{1}{L_\text{UV}},\nonumber\\
    \text{IR }(z=\infty)&:\ \  (e^{-f})'=\frac{1}{L_\text{IR}},\ \ (e^{-g})'=0,
\label{eq:UVIRasy}
\end{align}
For AdS$_{d+1}$ in the IR, we have
\begin{equation}
    a_{\mathrm{IR}}=\frac{L_{\mathrm{IR}}^{d-1}}{G_{d+1}}=\frac{e^{(D-d)g(\infty)}\mathrm{Vol}(M_{D-d})L_{\mathrm{IR}}^{d-1}}{G_N}=\frac{\mathrm{Vol}(M_{D-d})}{G_N}\left(e^{\frac{D-d}{d-1}g(\infty)}L_{\mathrm{IR}}\right)^{d-1},
\end{equation}
where $G_N$ is the $(D+1)$-dimensional Newton's constant, and $G_{d+1}$ is obtained by a standard Kaluza-Klein reduction with internal space metric $\hat g_{ij}=e^{2g(z)}g_{ij}$.  Taking $L_{\mathrm{IR}}\sim1/(e^{-f})'$, it is then natural to write down an unnormalized local holographic $c$-function of the form
\begin{equation}
    c(z)=\frac{1}{((e^{-\tilde f})')^{d-1}}.
\label{eq:twoDerivCfnct}
\end{equation}
In particular, the effective warp factor $\tilde f$ gives the precise combination of internal volume and AdS radius needed to obtain the IR central charge.  As before, one can verify that this is monotonic along flows
\begin{equation}
    c'(z)=-\frac{(d-1)(e^{-\tilde f})''}{((e^{-\tilde f})')^{d}}\le0,
\end{equation}
since $(e^{-\tilde f})''\ge0$ from NEC1, (\ref{eq:NEC1rew}).  Moreover, 

As before, we may define an effective AdS radius
\begin{equation}
    L_\text{eff}(z)=\frac{1}{(e^{-\tilde f})'},
\end{equation}
such that $L_\text{eff}'(z)\le 0$. The $c$-function is then simply
\begin{equation}
    c(z)=\frac{L_\text{eff}(z)^{d-1}}{G_N}.
\label{eq:2dcfad}
\end{equation}
Note, however, that $L_{\mathrm{eff}}$ defined here does not correspond directly to the radius of AdS$_{d+1}$; instead it is the AdS radius modified by the internal volume to account for the dimensionally reduced Newton's constant.  Moreover, unlike the domain wall flow case, this $c(z)$ diverges in the UV. This has a natural explanation: The $D$-dimensional theory appears to have an infinite number of $d$-dimensional degrees of freedom; \textit{i.e.}, as we approach the UV, the compact dimensions become large and we can no longer ignore the infinite KK tower of states. As it turns out, the divergence still encodes the UV central charge; we will return to this point in Section~\ref{sec:UVdiv}.

Note that NEC2, given in the form (\ref{eq:NEC2form}) also leads to a monotonicity of sorts.  In particular, as long as the internal curvature is non-positive, $\kappa\le0$, the quantity
\begin{equation}
    \mathcal{C}(z)=e^{(d-1)\tilde f}(f'-g'),
\end{equation}
satisfies the inequality
\begin{equation}
    \mathcal{C}'(z)\ge 0\qquad(\mbox{provided }\kappa\le0)
\end{equation}
Hence $\mathcal C(z)$ is a monotonically non-decreasing function towards the IR.  Moreover, making use of the IR behavior, (\ref{eq:UVIRasy}), we see that
\begin{equation}
    \mathcal{C}\overset{z\to\infty}{\sim}-\frac{e^{(D-d)g_\text{IR}}}{L_\text{IR}}\qty(\frac{L_\text{IR}}{z-z_0})^d<0,
\end{equation}
where $z_0$ is a constant offset. Since this is negative in the IR and the flow is non-decreasing towards the IR, we see that $\mathcal C(z)$ is negative along the entire flow.  Thus it must be the case that $f'<g'$ along the entire flow, so long as $\kappa\le 0$. It would be interesting to explore the implications of this condition as a second constraint on the flow (for $\kappa\le 0$).

\subsection{A concrete example: $\mathrm{AdS}_5\to\mathrm{AdS}_3$}

We now turn to four-derivative flows across dimensions where we include the Gauss-Bonnet coupling.  Since the expressions are somewhat lengthy for arbitrary UV and IR dimensions, $D$ and $d$, we start with a simple example of flowing from AdS$_5$ to AdS$_3\times T^2$ to motivate our procedure.  We thus take a metric of the form
\begin{equation}
    \dd{s}^2=e^{2f(z)}(-\dd{t}^2+\dd{x}^2+\dd{z}^2)+e^{2g(z)}(\dd{y}^2+\dd{w}^2).
\end{equation}
There are various explicit solutions in this class, including supergravity solutions describing flows of ${\cal N}=4$ SYM on $T^2$ \cite{Donos:2011pn,Almuhairi:2011ws,Benini:2015bwz,Uhlemann:2021itz}. The resulting NECs, in the presence of a Gauss-Bonnet term in the action, are obtained by orienting the null vectors along the $t$-$z$ and $t$-$y$ directions, respectively,
\begin{subequations}
    \begin{align}
        \text{NEC1: }&-\left(f''(z)-f'(z)^2\right)-2\left(g''(z)+g'(z)\left(g'(z)-2f'(z)\right)\right)\nonumber\\
        &+4\alpha e^{-2f(z)}g'(z)\Bigl[g'(z)\left(f''(z)-f'(z)^2\right)\nonumber\\
        &\kern3.0cm+2f'(z)\left(g''(z)+g'(z)\left(g'(z)-2f'(z)\right)\right)\Bigr]\ge0,\\
\text{NEC2: }&\left(f'(z)-g'(z)\right)'+\left(f'(z)-g'(z)\right)\left(f'(z)+2g'(z)\right)\nonumber\\
        &+4\alpha e^{-2f(z)}\Bigl[-\left(f'(z)g'(z)\left(f'(z)-g'(z)\right)\right)'\nonumber\\
        &\kern2.2cm+f'(z)g'(z)\left(f'(z)-g'(z)\right)\left(f'(z)-2g'(z)\right)\Bigr]\ge0.
    \end{align}
\label{eq:NEC12AdS53}%
\end{subequations}
These generalize the two-derivative NECs, \eqref{eq:twoDerivNEC}, in the case where $D=4$, $d=2$, and $\kappa=0$.  As a sanity check, note that NEC2 becomes trivial in the domain wall limit, $g=f$,  while NEC1 reduces to 
\begin{equation}
    -3\left(f''(z)-f'(z)^2\right) \left(1-4 \alpha e^{-2f} f'(z)^2\right)\ge0,
\end{equation}
in agreement with the domain wall flow case \eqref{eq:dwNEC}.

In order to obtain a $c$-function, note that, following (\ref{eq:NEC1rew}), the two-derivative NEC1 can be written as
\begin{equation}
    \left(e^{-\tilde f}\right)''\ge 6e^{-\tilde f}(g')^2,
\label{eq:AdS53two}
\end{equation}
where $\tilde f=f+2g$.  Examination of (\ref{eq:NEC12AdS53}) indicates that, in the presence of the Gauss-Bonnet correction, this can be extended to
\begin{equation}
    \qty(\qty(e^{-\tilde f})'+4\alpha e^{-\tilde f-2f}f'g'{}^2)'\ge 6e^{-\tilde f}(g')^2\qty[1+\frac{4}{3}\alpha\, e^{-2f}\qty((f'-g')^2-g'{}^2)].
\label{eq:AdS53four}
\end{equation}
Since $(g')^2$ is non-negative, the right-hand side of the two-derivative expression, (\ref{eq:AdS53two}), is non-negative.  However, the same cannot be said for (\ref{eq:AdS53four}), as the term inside the square brackets can in principle have either sign.  However, as long as we are working perturbatively in the higher derivative coupling, $\alpha$, this still leads to a monotonic expression for the left-hand side.

Validity of the perturbative expansion requires that the four-derivative Gauss-Bonnet term be parametrically smaller than the leading-order two derivative term, $\alpha R^2\ll R$, or $\alpha/\ell^2\ll1$ where $\ell$ is some radius of curvature of the background.  For the particular higher derivative flow at hand, (\ref{eq:AdS53four}), this corresponds to the two conditions
\begin{equation}
    \alpha e^{-2f}f'^2\ll1,\qquad \alpha e^{-2f}g'^2\ll1,
\label{eq:pertcond}
\end{equation}
in which case we can conclude that
\begin{equation}
    \qty(\qty(e^{-\tilde f})'+4\alpha e^{-\tilde f-2f}f'g'{}^2)'\ge0.
\label{eq:Lefficf}
\end{equation}
For a flow interpolating between the asymptotic regions given in (\ref{eq:UVIRasy}), we note that $e^{-f}\sim e^{-g}\sim z/L_{\mathrm{UV}}$ in the UV region, $z\to0$.  Then the perturbative conditions, (\ref{eq:pertcond}), translate into
\begin{equation}
   \frac{\alpha}{L_\text{UV}^2} \ll 1.
\end{equation}
While this changes along the flow, the first condition in (\ref{eq:pertcond}) corresponds to $\alpha/L_{\mathrm{eff}}\ll1$ where $L_{\mathrm{eff}}$ is an effective AdS radius interpolating between $L_{\mathrm{UV}}$ and $L_{\mathrm{IR}}$.  For the second condition in (\ref{eq:pertcond}), note that $e^{-g}$ interpolates from $z/L_{\mathrm{UV}}$ to a constant in the IR.  Hence $g'^2$ flows from $1/z^2$ to $0$.  Since $e^{-2f}$ scales as $z^2$ throughout the flow, the combination $e^{-2f}g'^2$ then interpolates between the values
\begin{eqnarray}
      e^{-2f} (g')^2&=&
    \begin{cases}
        0,&z\to \infty\text{ (IR)}\\
       \frac{1}{L_\text{UV}^2},& z\to 0\ \text{ (UV)}
    \end{cases}
\end{eqnarray}
The requirement that we are working perturbatively in $\alpha$ is, therefore,
\begin{equation}
   \left\{\frac{\alpha}{L_\text{UV}^2}, \frac{\alpha}{L_\text{IR}^2}\right\}\ll 1.\,
\end{equation}
at the endpoints of the flow, along with the assumption that the four-derivative corrections remain parametrically small along the flow.  This is equivalent to requiring that our EFT description remains valid.

With this in mind, one may generalize the two-derivative $c$-function defined in (\ref{eq:2dcfad}) by taking
\begin{equation}
    c(z)=\frac{L_{\mathrm{eff}}(z)}{G_N},
\label{eq:exCfnct}
\end{equation}
where now
\begin{equation}
    L_\text{eff}(z)=\frac{1}{\qty(e^{-\tilde{f}})'+4\alpha e^{-\tilde{f}-2f}f'(g')^2},
\label{eq:newLeff}
\end{equation}
is the Gauss-Bonnet corrected effective AdS$_3$ radius including the internal volume factor.  From (\ref{eq:Lefficf}), we immediately see that $L_{\mathrm{eff}}'(z)\le0$, so that $c'(z)\le0$.  As a result, $c(z)$ is monotonic non-increasing along the flow to the IR, so long as we work perturbatively in $\alpha$. Note that this $c$-function reduces to the two-derivative $c$-function in the IR where $g'=0$; this is a consequence of the fact that the Gauss-Bonnet term is trivial for AdS$_3$  and we might expect otherwise in general dimensions.

Turning our attention to NEC2, we see that it can be written as a total derivative
\begin{equation}
    \qty(e^{\tilde f}(f'-g')\qty(1-4\alpha e^{-2f}f'g'))'\ge0,
\end{equation}
which generalizes (\ref{eq:NEC2form}) for the case $\kappa=0$.  If we commit to being perturbatively small in $\alpha$, (\ref{eq:pertcond}), then the interpretation of NEC2 is almost identical to the two-derivative case \cite{GonzalezLezcano:2022mcd} as summarized above. We can define a function
\begin{equation}
    \mathcal{C}(z)=e^{\tilde f}(f'-g')\qty(1-4\alpha e^{-2f}f'g'),
\label{eq:calCfunc}
\end{equation}
such that $\mathcal{C}'(z)\ge 0$. In the IR, we have that
\begin{equation}
    \mathcal{C}(z)\overset{z\to\infty}{\sim}-\frac{e^{2g_\text{IR}}}{L_\text{IR}}\qty(\frac{L_\text{IR}}{z-z_0})^2<0,
\end{equation}
where $z_0$ is a constant. Since this is a negative in the IR and monotonically non-decreasing with respect to $z$, it must be the case that it is also negative in the UV. Hence, we have that $f'<g'$ along the entire flow.

\subsection{Gauss-Bonnet flows in arbitrary dimensions}
\label{sec:GBad}

Having examined flows from AdS$_5$ to AdS$_3$ we now turn to the general case of Gauss-Bonnet corrected flows in arbitrary dimensions.  Consider a flow from AdS$_{D+1}$ to AdS$_{d+1}$.  As noted above, we consider two conditions arising from the null energy condition, which we denoted NEC1 and NEC2.  Our main interest is in the $c$-function arising from NEC1, although NEC2 will also give rise to a monotonic function from the case $\kappa\le0$.

Making use of the curvature tensor components summarized in Appendix~\ref{app:Riemann}, we find the $t$-$z$ NEC1 to be given by
\begin{align}
    &-(d-1)(f''-(f')^2)-(D-d)(g''+g'(g'-2f'))\nonumber\\
    &+2\alpha e^{-2f}\Bigl[(d-1)(d-2)(f')^2\left((d-3)(f''-f'^2)+(D-d)(g''+g'(g'-2f'))\right)\nn\\
    &\kern2cm+2(D-d)(d-1)f'g'\left((d-2)(f''-f'^2)+(D-d-1)(g''+g'(g'-2f'))\right)\nn\\
    &\kern2cm+(D-d)(D-d-1)g'^2\left((d-1)(f''-f'^2)+(D-d-2)(g''+g'(g'-2f'))\right)\Bigr]\nonumber\\
    &-2\alpha \frac{\kappa}{\ell^2}(D-d)(D-d-1) e^{-2g}\Bigl[(d-1)(f''-f'^2)+(D-d-2)(g''+g'(g'-2f'))\Bigr]\ge0.
\label{eq:NEC1}
\end{align}
One can check that upon setting $f=g$ and $\kappa=0$, we get
\begin{align}
    (D-1)\qty((f')^2-f'')\qty(1-2\alpha(D-2)(D-3)e^{-2f} (f')^2)\ge0,
\end{align}
which is in perfect agreement with the domain wall flow NEC \eqref{eq:dwNEC}. As a sanity check, one can also see that setting $\alpha=0$ recovers the correct two-derivative result \eqref{eq:twoDerivNEC}.

We now seek a holographic $c$-function which could, {\it a priori}, be any arbitrary function
\begin{equation}
    c(z)=c(f,f',f'',...,g,g',g'',...;z).
\end{equation}
However, inspired by the form of the two-derivative $c$-function \eqref{eq:twoDerivCfnct} and the $\mathrm{AdS}_5\to\mathrm{AdS}_3$ case, namely (\ref{eq:exCfnct}) and (\ref{eq:newLeff}), a natural generalization would to be
\begin{equation}
c(z)=\frac{1}{((e^{-\tilde{f}})')^{d-1}}\to\frac{1+\mathcal O(\alpha)}{((e^{-\tilde{f}})'+\mathcal O(\alpha))^{d-1}},
\end{equation}
where the $\mathcal O(\alpha)$ terms are made from combinations of $f'$, $g'$ and $\kappa$.  Hence, we propose a candidate $c$-function
\begin{align}
    c(z)&=\frac{L_{\mathrm{eff}}(z)^{d-1}}{G_N}\Bigl[1+\alpha \qty(e^{-2f}\qty(a_1 (f')^2+a_2 f'g'+a_3 (g')^2)+b_1 e^{-2g}\frac{\kappa}{\ell^2})\Bigr],\nn\\
    L_{\mathrm{eff}}(z)&=\Bigl[\qty(e^{-\tilde{f}})'+\alpha\, e^{-\tilde{f}}\Bigl(e^{-2f}\qty(a_4 (f')^3+a_5 (f')^2g'+a_6 f'(g')^2+a_7 (g')^3)\nn\\
    &\kern3.5cm+e^{-2g}\frac{\kappa}{\ell^2}\qty(b_2f'+b_3g')\Bigr)\Bigr]^{-1},
\label{eq:LHcfnct}
\end{align}
for some choice of real coefficients $\{a_i, b_j\}$.  The structure of the central charge contains various occurring products of derivatives of the functions $f$ and $g$. Note that we  are interested in comparing to NEC1 in order to obtain monotonicity, and hence have avoided any terms with $f''$ or $g''$  in c(z) as these would lead to $f'''$ and $g'''$ terms in $c'(z)$, as well as $(f'')^2$ and $(g'')^2$ terms.

We now fix the coefficients $\{a_i,b_j\}$ by demanding monotonicity of $c(z)$, namely $c'(z)\le0$ under the assumptions of NEC1 and perturbative control.  To do so, we compute $c'(z)$ and adjust the coefficients to match the $f''$ and $g''$ terms with the structure of NEC1, namely (\ref{eq:NEC1}).  The expression for $c'(z)$ is not particularly illuminating, but it is given in Appendix~\ref{app:c'(z)} for completeness.  Comparing $c'(z)$ to NEC1, we see that for the particular choice of coefficients
\begin{align}
    a_1&=-2(d-1)(d-2),\nonumber\\
    a_2&=-4(D-d)(d-2),\nonumber\\
    a_3&=\mbox{arbitrary},\nn\\
    a_4&=0,\nonumber\\
    a_5&=\frac{4(D-d)(d-2)}{(d-1)},\nonumber\\
    a_6&=2(D-d)\qty(\frac{(2d-3)(D-d)}{(d-1)^2}-1)-\frac{a_3}{d-1},\nonumber\\
    a_7&=\mbox{arbitrary},\nn\\
    b_1&=2(D-d)(D-d-1),\nonumber\\
    b_2&=0,\nonumber\\
    b_3&=\frac{4(D-d)(D-d-1)}{(d-1)},
\label{eq:cfnctCoeffs}
\end{align}
we get monotonicity of the $c$-function, in the sense that
\begin{align}
    c'(z)= - \frac{e^{-\tilde{f}}(L_{\mathrm{eff}})^d}{G_N}\qty[\text{NEC1}+\frac{(D-1)(D-d)}{d-1}(g')^2\qty(1+\mathcal{O}(\alpha))]\le 0,\label{eq:ineq}
\end{align}
where we have made crucial use of the fact that we are working perturbatively in $\alpha$. Here, $\mathcal{O}(\alpha)$ denotes only terms which remain under perturbative control throughout the flow in the sense of (\ref{eq:pertcond}).  Notice also that \eqref{eq:LHcfnct} reduces to \eqref{eq:exCfnct} upon setting $D=4$, $d=2$, and $\kappa=0$, provided we take $a_3=a_7=0$.

Note that the two coefficients $a_3$ and $a_7$ are left undetermined; $a_7$ will be the coefficient of a term proportional to $(g')^2$ and so can never matter within the context of our analysis, and shifting $a_6$ is equivalent to a shift in $a_7$ and a shift in $a_3$ since  we are working perturbatively in $\alpha$.  This freedom in choosing $a_3$ and $a_7$ in principle yields a family of $c$-functions that all flow to the same IR central charge as $g'\to0$ in the IR.  However, the UV behavior will be affected, and below we will find a preferred combination of these coefficients.  In fact, if one were to relax the above condition \eqref{eq:ineq} by replacing NEC1 with NEC1$\times(1+\mathcal{O}(\alpha))$, then it would become apparent that, due to the perturbative nature of our analysis, there are actually five free parameters rather than the na\"ively apparent two.  Intuitively, this is equivalent to the freedom of perturbatively combining the numerator of \eqref{eq:LHcfnct} with its denominator.  It is convenient, however, to keep these terms separate when taking the IR limit, as we will see in Section.~\ref{sec:IR}.

We may also consider NEC2, which can be arranged in the form
\begin{align}
    &\Big\{e^{(d-1)\tilde{f}}\Big[(f'-g')+2\alpha\Bigl(e^{-2f}(f'-g')\Big(-(d-1)(d-2)(f')^2\nn\\
    &\kern3cm-2(d-1)(D-d-1)f'g'-(D-d-1)(D-d-2)(g')^2\Big)\nonumber\\
    &\kern3cm+e^{-2g}\frac{\kappa}{\ell^2}\Big((D-4d-1)f'+(-5+(8-3D)D+d(-8+6D))g'\Big)\Bigr)\Big]\Big\}'\ge\nonumber\\
    &-\fft\kappa{\ell^2}e^{(d-1)\tilde{f}+2f-2g}\Bigl[D-d-1+2\alpha\Bigl(e^{-2f}\Bigl(3d(d+1)f''+3d^2g''\nn\\
    &\kern4cm+(D+2d+2)d(d-1)(f')^2-d(D(3-2D)+dD+4d^2+2)f'g'\nonumber\\
    &\kern4cm+(D-d-2)(D^2-2dD-4D-2(d-2)+3)(g')^2\Bigr)\nonumber\\
    &\kern4cm+\fft\kappa{2\ell^2}(D-d-1)(7+2d^2-4d(D-2)+2D(D-4))\Bigr)\Bigr].
\end{align}
One may check that setting $g=f$ and $\kappa=0$ makes the left- and right-hand sides of this inequality identically zero, as it should.  This suggests that we define
\begin{align}
    \mathcal{C}(z)=&e^{(d-1)\tilde{f}}\Big[(f'-g')+2\alpha\Bigl(e^{-2f}(f'-g')\Big(-(d-1)(d-2)(f')^2\nn\\
    &\kern2cm-2(d-1)(D-d-1)f'g'-(D-d-1)(D-d-2)(g')^2\Big)\nonumber\\
    &\kern2cm+e^{-2g}\frac{\kappa}{\ell^2}\Big((D-4d-1)f'+(-5+(8-3D)D+d(-8+6D))g'\Big)\Bigr)\Big],
\end{align}
analogous to (\ref{eq:calCfunc}) for the case of $\mathrm{AdS}_5\to\mathrm{AdS}_3$.  NEC2 is then the statement that
\begin{equation}
    \mathcal C'(z)\ge-\frac\kappa{\ell^2}e^{(d-1)\tilde{f}+2f-2g}(D-d-1)\qty(1+\mathcal{O}(\alpha)).
\end{equation}
Then $\mathcal{C}'(z)>0$ for $\kappa=-1$ and $\mathcal{C}'(z)\ge0$ for $\kappa=0$, so long as the $\mathcal O(\alpha)$ corrections are parametrically small. Then in the IR, we find that
\begin{equation}
    \mathcal{C}(z)\overset{z\to\infty}{\sim}-\frac{e^{(D-d)g_\text{IR}}}{L_\text{IR}}\qty(\frac{L_\text{IR}}{z-z_0})^d\qty(1-\frac{2\alpha(d-1)(d-2)}{L_\text{IR}^2}+\alpha e^{-2g_\text{IR}}\frac{\kappa}{\ell^2}(D-4d-1))<0.
\end{equation}
This should hold so long as $\alpha/L_\text{IR}^2\ll 1$ and $\alpha/\ell^2\ll 1$. Then since $\mathcal{C}(z)$ is negative in the IR and non-decreasing as $z$ increases, we conclude that it must always be negative. This imposes a constraint
\begin{align}
    0>\,&(f'-g')+2\alpha\Bigl(e^{-2f}(f'-g')\Big(-(d-1)(d-2)(f')^2\nn\\
    &\kern2cm-2(d-1)(D-d-1)f'g'-(D-d-1)(D-d-2)(g')^2\Big)\nonumber\\
    &\kern2cm+e^{-2g}\frac{\kappa}{\ell^2}\Big((D-4d-1)f'+(-5+(8-3D)D+d(-8+6D))g'\Big)\Bigr).
\end{align}
Heuristically, this provides an additional constraint to the $c$-function considerations.

\subsubsection{Generic Einstein internal manifolds}\label{sec:generalization}
If we relax the condition that the internal manifold is maximally symmetric, and instead only require it to be an Einstein manifold with Ricci curvature
\begin{equation}
    \tilde{R}_{ij}=k g_{ij},
\end{equation}
with $g$ the metric on the internal space, then the null energy condition will be, in general, more complicated. In particular, we no longer know the internal Riemann tensor $\tilde{R}_{ijkl}$; however, the only component of the full Riemann tensor $\hat{R}_{MNPQ}$ that contains the uncontracted internal Riemann tensor is $\hat{R}_{ijkl}$ with all internal indices, which will not affect the $t$-$z$ null energy condition NEC1. Then all the previous arguments hold for the monotonicity of the $c$-function if we replace
\begin{equation}
    \frac{\kappa}{\ell^2}\to\frac{k}{D-d-1}.
\end{equation}
However, the same is not true of NEC2 since it would be dependent on $\tilde{R}_{ijkl}$ in general.

\subsection{Changing Coordinates}

While we have parametrized the bulk metric according to (\ref{Eq:Flowfg}), in some situations it is convenient for one to work in a different gauge,
\begin{equation}
    \dd{s}^2=e^{2A(r)}(-\dd{t}^2+\dd{\vec{x}}^2)+\dd{r}^2+e^{2B(r)}\dd{s}_{M_{D-d}}^2.\label{eq:AltCoords}
\end{equation}
The $c$-functions we have defined do not depend on the choice of coordinates.  Nevertheless, we present NEC1 and the corresponding $c(r)$ function in Appendix~\ref{app:altCoords1} in case such expressions may prove useful.

\section{Fixed point limits of the $c$-function}\label{sec:limits}

In Section~\ref{sec:LH}, we have constructed a monotonic $c$-function, (\ref{eq:LHcfnct}) with coefficients given in (\ref{eq:cfnctCoeffs}), for Einstein-Gauss-Bonnet flows across dimensions.  This $c$-function is a natural extension of its two-derivative counterpart, (\ref{eq:twoDerivCfnct}), as well as the higher-derivative $c$-function, (\ref{eq:dwcfnct}), for flows in the same dimension.  To better understand the physics of this NEC-motivated $c$-function, we now consider its UV and IR limits and compare it to the expected central charges at the endpoints of the flow.

\subsection{The IR limit}\label{sec:IR}
One important reason for considering higher derivatives is that they break the degeneracy between the $a$-type and $c$-type central charges. Focusing on $d=4$ for the moment, the Gauss-Bonnet correction splits the two central charges in the IR \cite{Myers:2010xs}
\begin{subequations}
    \begin{align}
        a&=\frac{L_\text{IR}^3}{G_5}\qty(1-\frac{12\alpha}{L_\text{IR}^2}),\\
        c&=\frac{L_\text{IR}^3}{G_5}\qty(1-\frac{4\alpha}{L_\text{IR}^2}).
    \end{align}
\label{eq:4DIRcc}%
\end{subequations}
If we do not include higher derivatives, then these are exactly the same. In particular, a holographic two-derivative flow cannot tell the difference between whether $a$ or $c$ is flowing monotonically.  However, for the four-derivative central charge, (\ref{eq:LHcfnct}), we find the IR limit
\begin{equation}
    c(z)\overset{z\to\infty}{\sim}\fft{L_\text{IR}^{d-1}}{G_{d+1}}\qty(1-\frac{2\alpha(d-1)(d-2)}{L_\text{IR}^2}),
\label{eq:IRlimit}
\end{equation}
where, as we show below, the $(d+1)$-dimensional Newton's constant is
\begin{equation}
\fft1{G_{d+1}}=
\frac{e^{(D-d)g_\text{IR}}\mathrm{Vol}(M_{D-d})}{G_N}\left(1+2\alpha(D-d)(D-d-1)\frac{\kappa}{\ell^2}e^{-2g_\text{IR}}\right).
\end{equation}

While the above holds for arbitrary $D$ and $d$, we can compare with the four-dimensional IR central charges, (\ref{eq:4DIRcc}), by setting $d=4$.  In this case, we get that
\begin{equation}
    c(z)\overset{z\to\infty}{\sim}\fft{L_\text{IR}^3}{G_5}\qty(1-\frac{12\alpha}{L_\text{IR}^2}),
\end{equation}
with
\begin{equation}
    \fft1{G_5}=
\frac{e^{(D-4)g_\text{IR}}}{G_N}\left(1+2\alpha(D-4)(D-5)\frac{\kappa}{\ell^2}e^{-2g_\text{IR}}\right).
\label{eq:GNren}
\end{equation}
This then clearly reduces to the $a$ central charge as we would expect from the $a$-theorem, and notably is not the $c$ central charge.

More generally, we expect the $A$-type central charge in the IR to be \cite{Imbimbo:1999bj,Hung:2011xb}
\begin{equation}
    A=\frac{L_\text{IR}^{d-1}}{G_{d+1}}\qty(1-2(d-1)(d-2)\frac{\alpha}{L_\text{IR}^2}),
\end{equation}
which precisely matches the IR limit, \eqref{eq:IRlimit}. Hence, the $c$-function originating from NEC1 pertains to the monotonicity of what becomes the $A$-type central charge in the IR. Note that we have not imposed this fact; simply solving for the allowed parameters $\{a_i,b_i\}$ in (\ref{eq:LHcfnct}) that give monotonicity from NEC1 has demanded that the IR limit be unambiguously the $A$-type central charge.

We now return to the relation between the $(D+1)$-dimensional and $(d+1)$-dimensional Newton's constant, (\ref{eq:GNren}).  The lower-dimensional Newton's constant is obtained from compactification of the gravitational part of the Lagrangian, (\ref{eq:Lgrav}).  In the IR, the spacetime is AdS$_{d+1}\times M_{D-d}$.  Furthermore, in this limit, the Gauss-Bonnet term, (\ref{eq:chi4}), splits as
\begin{equation}
    \chi_4\to\chi_4+\tilde\chi_4+2R\tilde R,
\end{equation}
where $\tilde{R}$ is the internal Ricci scalar and $\tilde{\chi}_4$ is the internal Gauss-Bonnet term
\begin{align}
    \tilde{R}&=\frac{\kappa}{\ell^2}(D-d)(D-d-1),\nn\\
    \tilde{\chi}_4&=\frac{1}{\ell^4}(D-d)(D-d-1)(D-d-2)(D-d-3).
\label{eq:tRtchi}
\end{align}
Then the gravitational action reduces as
\begin{align}
    S&=\fft1{16\pi G_N}\int d^{D+1}x\sqrt{-g}\qty[R+\fft{D(D-1)}{L^2}+\alpha\chi_4]\nn\\
    &=\fft1{16\pi G_N}\int d^{d+1}x\sqrt{-g_{d+1}}\int d^{D-d}ye^{(D-d)g_{IR}}\sqrt{g_{D-d}}\nn\\
    &\kern3cm\times\biggl[\left(1+2\alpha\tilde{R}\right)R+\fft{D(D-1)}{L^2}+\alpha\chi_4+\tilde R+\alpha\tilde{\chi}_4\biggr].
\end{align}
Integrating out the internal coordinates gives the $(d+1)$-dimensional Newton's constant
\begin{equation}
    \fft1{G_{d+1}}=\fft{e^{(D-d)g_{IR}}\mathrm{Vol}(M_{D-d})}{G_N}\left(1+2\alpha\tilde{R}\right),
\end{equation}
where $\tilde R$ is given in (\ref{eq:tRtchi}).  Making this substitution for $\tilde R$ then yields the expression given above in (\ref{eq:GNren}).

\subsection{The UV divergence}\label{sec:UVdiv}

We now turn to the UV behavior of the $c$-function, (\ref{eq:LHcfnct}).  As is often the case when defining $c$-functions in flows across dimensions, this function diverges in the UV.  This, of course, is not surprising since we will see an infinite number of lower-dimensional degrees of freedom in the UV. While (\ref{eq:LHcfnct}) does not interpolate between the UV and IR central charges, we can still ask whether its UV divergence can be related to the UV central charge.  To answer this question, we first look at the two-derivative case.

\subsubsection{Two-derivative case}
Ignoring higher-derivative corrections for the moment, for general $D$ to $d$ dimensional flows, the $c$-function is given by (\ref{eq:twoDerivCfnct}), which we write out as
\begin{equation}
    c(z) = e^{(d-1)f+(D-d)g}\left(-f' - \frac{D-d}{d-1}g'\right)^{-(d-1)}.
\end{equation}
In the UV we have $e^f\sim e^g\sim L_\text{UV}/z$, so
\begin{equation}
    c(z) \overset{z\to 0}{\sim} \left(\frac{L_\text{UV}}{z}\right)^{D-1}\left(\frac{d-1}{D-1}z\right)^{d-1} ~\propto \quad \frac{(L_\text{UV})^{D-1}}{z^{D-d}}.
\end{equation}
The numerator gives the unnormalized UV central charge. The denominator diverges with increasing energy scale, and the power is the number of compact dimensions.

\subsubsection{Gauss-Bonnet}
Now we consider what happens when we reintroduce the Gauss-Bonnet term. For the case of no internal curvature, $\kappa=0$, the $c$-function in (\ref{eq:LHcfnct}) reduces to
\begin{equation} \label{k=0 c func}
    c(z)=e^{(d-1)f+(D-d)g}\frac{1+\alpha e^{-2f}\qty(a_1 (f')^2+a_2f'g'+a_3(g')^2)}{\qty(-f'-\frac{D-d}{d-1}g'+\alpha e^{-2f}\qty(a_5(f')^2g'+a_6f'(g')^2+a_7(g')^3))^{d-1}}\;,
\end{equation}
which, in the UV limit, behaves as 
\begin{align}
    c(z)&\overset{z\to 0}{\sim}\left(\frac{L_\text{UV}}{z}\right)^{D-1}\frac{1+\frac{\alpha}{{L_\text{UV}^2}}\qty(a_1 +a_2+a_3)}{\qty(\frac{D-1}{d-1}\frac{1}{z} - \frac{\alpha}{L_\text{UV}^2}\qty(a_5+a_6+a_7)\frac{1}{z})^{d-1}} \nonumber\\
    &= \qty(\frac{d-1}{D-1})^{d-1}\frac{(L_\text{UV})^{D-1}}{z^{D-d}} \frac{1+\frac{\alpha}{{L_\text{UV}^2}}\qty(a_1 +a_2+a_3)}{\qty(1-\frac{\alpha}{{L_\text{UV}^2}}\frac{d-1}{D-1}\qty(a_5+a_6+a_7))^{d-1}} \label{4 deriv UV}\;.
\end{align}
Note that the curvature terms proportional to $\kappa/\ell^2$ do not affect the UV limit \eqref{4 deriv UV} since $e^{-2g}f'\sim z$ and $e^{-2g}\sim z^2$ in the UV, which is to be expected since intuitively the ``compact'' dimensions will appear large at very high energies. If we demand that $c(z) \propto a_\text{UV}/z^{D-d}$ in the UV limit, (\ref{4 deriv UV}) places constraints on sums of the $a$ coefficients. In particular, comparing to the known result,
\begin{equation}
    a_\text{UV}=\frac{L_\text{UV}^{D-1}}{G_N}\qty(1-2(D-1)(D-2)\frac{\alpha}{L_\text{UV}^2}),
\end{equation}
we must satisfy
\begin{equation}
    a_1+a_2+a_3+\frac{(d-1)^2}{D-1}(a_5+a_6+a_7)=-2(D-1)(D-2),\label{eqn:UVacons}
\end{equation}
which corresponds to the requirement that
{\footnotesize
\begin{align}
    a_7=&\frac{D-1}{(d-1)^2}\Bigg[2\frac{(D-1)((D+1)(D-4)-3d(d-3))+(d-3)(d-1)(D-d)+(2d-3)(D-d)^2}{(D-1)}-\frac{D-d}{d-1}a_3\Bigg].
\label{eqn:a7req}
\end{align}}%
This provides an additional constraint on the coefficients \eqref{eq:cfnctCoeffs}, reducing the number of free coefficients from two to one. Given the discussion hitherto, we may always impose this additional requirement.

\section{Higher-derivative gravity and holographic entanglement entropy}\label{sec:EE}

In this section, we discuss the construction of monotonic $c$-functions from the perspective of holographic entanglement entropy. It is well-known that finding the entanglement entropy of a region in a holographic CFT is equivalent to finding a bulk surface minimizing some choice of functional; at the two-derivative level, this is just the Ryu-Takayanagi (RT) area functional \cite{Ryu:2006bv,Ryu:2006ef}. However, minimizing the area of the extremal surface is insufficient when higher derivatives are present; in particular it has been argued  \cite{Faulkner:2013ana,Hung:2011xb,deBoer:2011wk,Bhattacharyya:2013jma} that, given a theory described by the Einstein-Gauss-Bonnet action
\begin{eqnarray}
    I_{\rm total}&=& I_{\rm bulk} + I_{\rm GH} + I_{\rm ct}, \nonumber \\
    I_{\rm bulk}&=& \int\dd[D+1]{x}\qty[R+\frac{D(D-1)}{L^2}+\alpha\chi_4], \nonumber \\
    I_{\rm GH}&=&\int\dd[D]{x}\qty[ K -2\alpha \qty(G_{ab}K^{ab}+\frac{1}{3}\qty(K^3-3KK_2 +2K_3))],
\end{eqnarray}
where $K_{ab}$ is the extrinsic curvature with trace $K$, $K_2=(K_{ab})^2$, and $K_3=K_{ab}K^{bc}K_{c}^{\ \, a}$, the RT functional must be replaced with the Jacobson-Myers (JM) functional\footnote{Note that for black holes, the Jacobson-Myers functional leads to the same result as Wald's entropy \cite{Wald:1993nt,Iyer:1994ys,Jacobson:1993vj}, but it is generically different.} \cite{Jacobson:1993xs}
\begin{equation}
    S_\text{JM}=\frac{1}{4G_N}\int_\Sigma \dd[d] x \sqrt{h}\left(1+2\alpha {\cal R}\right)+\frac{1}{2G_N}\int_{\partial\Sigma}\dd[d-1]{x}\sqrt{\tilde h}2\alpha{\cal K},
\end{equation}
where $\Sigma$ is the surface over which the functional is being minimized with boundary $\partial\Sigma$, $h$ is the determinant of the induced metric on $\Sigma$, $\tilde h$ is the induced metric (of the induced metric $h$) on $\partial\Sigma$, ${\cal R}$ is the scalar curvature of $\Sigma$, and $\cal K$ is the trace of the extrinsic curvature of the boundary $\partial\Sigma$. The term containing $\cal{K}$ may be viewed as a Gibbons-Hawking term that renders the variational principle well-defined. The equation  of motion that follows from the JM functional is
\begin{equation}
    {\cal K}+2\alpha ({\cal R}{\cal K}-2{\cal R}_{ij}{\cal K}^{ij})=0.
\end{equation}
We may then compute the holographic entanglement entropy of a region $A$ by minimizing this functional over all surfaces homologous to $A$
\begin{equation}
    S_\text{EE}=\min_{\Sigma\sim A}S_\text{JM}(\Sigma).
\end{equation}

The goal of this section is to construct monotonic $c$-functions from the entanglement entropy. For flows down to AdS$_3$, it is natural to obtain a monotonic $c$-function as the coefficient of the logarithmic term \cite{Casini:2006es}
\begin{equation} \label{eqn:c from S}
    c_\text{EE}=R\partial_R S_\text{EE},
\end{equation}
where $R$ is the radius of the entangling region. An analogous quantity that interpolates between free energies in AdS$_4$ flows is
\begin{equation}\label{eqn:F from S}
    c_\text{EE} = R\partial_R S_\text{EE} - S_\text{EE}\;,
\end{equation}
and its monotonicity can be proven using strong subadditivity on field-theoretic grounds \cite{Casini:2017vbe}. However, it is not clear how to define similar quantities for AdS$_5$ and above. Strong subadditivity may be used to construct monotonic functions in higher dimensions, but they no longer interpolate between central charges at the fixed points.

\subsection{AdS$_{D+1}\to$ AdS$_3$}\label{sec:EEcfnct}
It is most tractable to look at flows from AdS$_{D+1}$ down to AdS$_3$. Equivalently, this may be viewed as a flow from CFT$_D$ to CFT$_2$. Generically, we have a metric of the form \eqref{Eq:Flowfg}, but we will further specify the metric to be
\begin{equation}
    \dd s^2=e^{2f(z)}\qty(-\dd t+\dd z^2+\dd r^2)+e^{2g(z)}\dd s_{M_{D-2}}^2,
\end{equation}
with asymptotic behavior
\begin{align}
    &z\to0:\ &f(z)\to\log\qty(L_\text{UV}/z),&\qquad g(z)\to\log\qty(L_\text{UV}/z),\nonumber\\
    &z\to\infty:\ &f(z)\to\log\qty(L_\text{IR}/z),&\qquad g(z)\to g_\text{IR}.
\end{align}
Our CFT$_D$ lives on $\mathbb{R}^{1,1}\times M_{D-2}$. We will consider entangling regions%
\footnote{For a more detailed discussion of choices of entangling regions in flows across dimensions, see \cite{GonzalezLezcano:2022mcd}.}
which wrap the internal $M_{D-2}$.  The induced metric on a constant time slice parameterized by a profile $r(z)$ is
\begin{equation}
    \dd{\sigma}^2=e^{2f}(1+r'(z)^2)\dd{z}^2+e^{2g}\dd{s}^2_{M_{D-2}}.
\end{equation}
We will assume boundary conditions
\begin{equation}
    r(0)=R,\qquad r(z_0)=0,\qquad r'(z_0)=-\infty, 
\end{equation}
where $R$ is the radius of the entangling region and $z_0$ is the deepest point in the bulk that the minimal surface probes along the holographic radial coordinate, that is, the turning point of the surface in the mechanical analogy. In terms of this profile, the induced Ricci scalar is
\begin{align}
    \mathcal{R}=&(D-2)(D-3)\frac{\kappa}{\ell^2}e^{-2g}\nonumber\\
    &+(D-2)\frac{e^{-2f}}{\left(1+(r')^2\right)^2}\Bigg[\qty(1+(r')^2)\qty(2f'g'-(D-1)(g')^2-2g'')+2g'r'r''\Bigg],\label{eq:Rind}
\end{align}
which, after some integration by parts, leads to a JM functional 
\begin{align} \label{eqn:AdS3JM}
    S_\text{JM}&=\frac{2\text{Vol}\qty(M_{D-2})}{4G_N}\int\dd{z}e^{\tilde f}\qty[\sqrt{1+(r')^2}\qty(1+2\tilde\alpha\frac{\kappa}{\ell^2}e^{-2g})+2\tilde\alpha \frac{e^{-2f}(g')^2}{\sqrt{1+(r')^2}}],
\end{align}
where the rescaled Gauss-Bonnet coupling
\begin{equation}
    \tilde\alpha\equiv\alpha(D-2)(D-3),
\end{equation}
is introduced for convenience.  Here we have ignored the boundary term from integrating by parts since it will automatically cancel with the Gibbons-Hawking term $\cal K$. Since this functional is independent of $r(z)$, $S_\text{JM}$ admits a first integral
\begin{equation}
    C=\frac{r' e^{\tilde f} \left(\left((r')^2+1\right) \left(1+2 \tilde\alpha \frac{\kappa}{\ell^2}e^{-2g} \right)-2\tilde\alpha e^{-2f} (g')^2\right)}{\left((r')^2+1\right)^{3/2}},
\end{equation}
which can be solved to give
\begin{equation}
    r'(z)=-\frac{\mathcal{F}}{\sqrt{1-\mathcal{F}^2+4\tilde\alpha\qty(\frac{\kappa}{\ell^2}e^{-2g}-e^{-2f}(g')^2(1-\mathcal{F}^2))}},\ \ \mathcal{F}(r)\equiv C e^{-\tilde f},\label{eq:rprime}
\end{equation}
or, equivalently,
\begin{equation}
    z'(r)=-\frac{\sqrt{1-\mathcal{F}^2+4\tilde\alpha\qty(\frac{\kappa}{\ell^2}e^{-2g}-e^{-2f}(g')^2(1-\mathcal{F}^2))}}{\mathcal{F}}.\label{eq:zprime}
\end{equation}
To fix the value of $C$, we note that we should have $r'(z)\to-\infty$ as $z\to z_0$; this then requires that
\begin{equation}
    C=e^{\tilde f_0}\qty(1+2\tilde\alpha\frac{\kappa}{\ell^2}e^{-2g_0})\quad\text{where}\quad \tilde f_0=\tilde f(z_0),\ g_0=g(z_0).
\end{equation}

Recall that we are interested in obtaining a monotonic $c$-function from the entanglement entropy following \eqref{eqn:c from S}, where $R$ is given by
\begin{align}
    R=&-\int_0^{z_0}\dd{z}r'(z).
\end{align}
The negative sign is due to the fact that $r'(z)$ is negative in this parameterization. We know $r'(z)$ from the integral of motion, \eqref{eq:rprime}, and so we may write
\begin{align}
    R=&\int_0^{z_0}\dd{z}\frac{\mathcal{F}}{\sqrt{1-\mathcal{F}^2+4\tilde\alpha\qty(\frac{\kappa}{\ell^2}e^{-2g}-e^{-2f}(g')^2(1-\mathcal{F}^2))}}\nonumber\\
    =&\int_0^{z_0}\dd{z}\qty[\frac{\mathcal{F}}{\sqrt{1-\mathcal{F}^2+4\tilde\alpha\frac{\kappa}{\ell^2}e^{-2g}}}+2\tilde\alpha \frac{e^{-2f}(g')^2\mathcal{F}}{\sqrt{1-\mathcal{F}^2}}]+\mathcal{O}(\tilde\alpha^2).
\end{align}
Note that in the second line, we have partially expanded the denominator; this will be important to avoid triple derivatives from integrating by parts. As in the two-derivative case, the integrand is divergent at the cap-off point $z_0$, so it must be integrated by parts to give
\begin{align}
    R=&\lim_{\epsilon\to0}\int_{r_0}^{r_c}\dd{r}\qty[\sqrt{1-\mathcal{F}^2+4\tilde\alpha\frac{\kappa}{\ell^2}e^{-2g}}\dv{}{r}\frac{1}{\mathcal{F}'+\frac{4\tilde\alpha}{\mathcal{F}}\frac{\kappa}{\ell^2}e^{-2g}g'}+2\tilde\alpha \sqrt{1-\mathcal{F}^2}\dv{}{r}\qty(\frac{e^{-2f}(g')^2}{\mathcal{F}'})]\nonumber\\
    &+2\tilde\alpha\lim_{\epsilon\to0}\frac{e^{-2f}(g')^2}{\mathcal{F}'}\Bigg\vert_{z=\epsilon}+\mathcal{O}(\tilde\alpha^2).\label{eq:R}
\end{align}
The profile $r(z)$ has been useful for obtaining an expression for $R$, but it will now be useful to phrase matters in terms a profile $z(r)$ with boundary conditions
\begin{equation}
    z(0)=z_0,\ \ z'(0)=0,\ \ z(R)=0.
\end{equation}
The induced Ricci scalar with respect to this profile is
\begin{align}
    \mathcal{R}=&(D-2)(D-3)\frac{\kappa}{\ell^2}e^{-2g}\nonumber\\
    &+(D-2)\frac{e^{-2f}(z')^2}{\left(1+(z')^2\right)^2}\Bigg[\qty(1+(z')^2)\qty(2f'g'-(D-1)(g')^2-2g'')+2g'z'z''\Bigg],
\end{align}
which leads to a Jacobson-Myers functional of the form
\begin{align}
    S_\text{JM}=&\frac{2\text{Vol}(M_{D-2})}{4G_N}\int_0^{R_c}\dd{r}e^{\tilde f(z(r))}\Bigg[\sqrt{1+(z')^2}\qty(1+2\tilde\alpha\frac{\kappa}{\ell^2}e^{-2g(z(r))})\nonumber\\
    &\kern5cm+2\tilde\alpha\frac{e^{-2f(z(r))}(g')^2(z')^2}{\sqrt{1+(z')^2}}\Bigg]-2\tilde\alpha e^{\tilde f-2f}g'\Big\vert_{r=R_c},\label{eq:SJM}
\end{align}
where $R_c$ is the cutoff value of $R$ such that $z(R_c)=\epsilon$. The boundary term, while divergent, is independent of $R$ and so will not cause us any issues. Since the UV boundary condition has the form $z_R(r=R_c)=\epsilon$, varying this boundary condition with respect to $R$ gives the relation
\begin{equation}
    z'\dv{R_c}{R}+\dv{z}{R}=0.\label{id1}
\end{equation}
Moreover, as $\epsilon\to 0$, $\dd R_c/\dd R\to 1$ at the boundary. One may now apply $R\partial_R$ to \eqref{eq:SJM} and impose the equations of motion. Using the relation \eqref{id1}, the monotonic central charge is then given by
\begin{equation}
    c_\text{EE}=\frac{2\text{Vol}(M_{D-2})}{4G_N}e^{\tilde f_0}\qty(1+2\tilde\alpha\frac{\kappa}{\ell^2}e^{-2g_0})R.\label{eq:EEcfnct}
\end{equation}
This generalizes the two-derivative case \cite{GonzalezLezcano:2022mcd} by simply using the four-derivative first integral $C$ rather than the two-derivative one $e^{\tilde{f}_0}$. Using the identity
\begin{equation}
    \pdv{\mathcal{F}}{z_0}=\qty(\tilde f'(z_0)-4\tilde\alpha\frac{\kappa}{\ell^2}e^{-2g_0}g'(z_0))\mathcal{F},
\end{equation}
substituting our expression for $R$ \eqref{eq:R} into $c_\text{EE}$, and differentiating with respect to $z_0$, we can show that
\begin{align}
    \dv{c_\text{EE}}{z_0}=&\frac{2\text{Vol}\qty(M_{D-2})}{4G_N}\int_0^{R}\dd r\frac{e^{\tilde f}\mathcal{F}^2\tilde f'_0}{\sqrt{1-\mathcal{F}^2}(\tilde f')^2}\Bigg\{(\tilde f')^2-\tilde f''+\tilde\alpha\frac{\kappa}{\ell^2}\frac{e^{-2g_0}g_0'}{\tilde f_0'}\qty((\tilde f')^2-\tilde f'')\nonumber\\
    &+\tilde\alpha e^{-2f}\qty[-2f''(g')^2+4g''f'g'+2(D-2)g''(g')^2-2(f')^2(g')^2+2(D-2)(g')^4]\nonumber\\
    &+2\tilde\alpha\frac{\kappa}{\ell^2}\frac{e^{2(\tilde f-2\tilde f_0-g)}}{\qty(e^{2\tilde f_0}-e^{2\tilde f})\tilde f'}\Bigg[-4e^{4\tilde f_0}(f')^3+3\qty(4e^{\tilde f}-(D-4)e^{4\tilde f_0}-6e^{2(\tilde f+\tilde f_0)})(f')^2g'\nonumber\\
    &-(D-2)\qty(-4(3D-8)e^{\tilde f}+(D(D+10)+20)e^{4\tilde f_0}+6(3D-8)e^{2(\tilde f+\tilde f_0)})(g')^3\nonumber\\
    &+\qty(-8e^{\tilde f}+(D-6)e^{4\tilde f_0}+12e^{2(\tilde f+\tilde f_0)})g'f''+(D-2)\qty(-4e^{\tilde f}+(D-4)e^{4\tilde f_0}+6e^{2(\tilde f+\tilde f_0)})g'g''\nonumber\\
    &+\qty(2(3D-7)(g')^2+g'')f'\qty(4e^{\tilde f}-6e^{2(\tilde f+\tilde f_0)})\nonumber\\
    &+4e^{4\tilde f_0}f'\qty(-(3D(D-8)+40)(g')^2+f''+Dg'')\Bigg]\Bigg\},\label{eq:dcmono/dz0}
\end{align}
where, for notational simplicity, we have denoted
\begin{equation}
    \tilde f_0'=\tilde f'(z_0),\ \ g_0'=g'(z_0).
\end{equation}
The above formula \eqref{eq:dcmono/dz0} of course assumes the use of the integral of motion \eqref{eq:zprime}. Note that this agrees with \cite{GonzalezLezcano:2022mcd} for $\tilde\alpha=0$.

If one sets $\kappa=0$, then we see that, schematically,
\begin{equation}
    \dv{c_\text{EE}}{z_0}=-\frac{2\text{Vol}\qty(M_{D-2})}{4G_N}\int_0^{R}\dd r\frac{e^{\tilde f}\mathcal{F}^2\tilde f'(z_0)}{\sqrt{1-\mathcal{F}^2}(\tilde f')^2}\qty[\text{NEC1}+(D-1)(D-2)(g')^2\qty(1+\mathcal{O}\qty(\tilde\alpha))]\le 0.
\end{equation}
Thus, for $\kappa=0$, we recover a notion of monotonicity along flows from the UV to the IR. Unfortunately, for $\kappa\ne0$, it is unclear what to make of the resulting expression. 

Note that upon setting $D=2$, $g=f$, and $\kappa=0$, one recovers the result for the strip in flows within the same dimension \cite{Myers:2012ed}. The comparison is more direct in the coordinates \eqref{eq:AltCoords}; the expression \eqref{eq:dcmono/dz0} is reexpressed in said coordinates in Appendix \ref{app:altCoords2}.

\subsection{Relation to the NEC-motivated $c$-function}
It is interesting to note that the monotonic $c$-function \eqref{eq:EEcfnct} constructed from the entanglement entropy is in fact related to the NEC-motivated $c$-function \eqref{eq:LHcfnct}, at least for $\kappa=0$. For arbitrary $D$ with $\kappa=0$, we have 
\begin{equation}
    R = \int_0^{z_0}\dd z\frac{\mathcal{F}}{\sqrt{1-\mathcal{F}^2}}\left(1+2\tilde\alpha e^{-2f}(g')^2\right) \; ,
\end{equation}
and the entropic $c$-function is
\begin{equation}
    c_{\text{EE}}(z_0) \propto e^{f_0+(D-2)g_0}R
    = \int_0^{z_0}\dd z\frac{e^{f + (D-2) g}\mathcal{F}^2}{\sqrt{1-\mathcal{F}^2}}\left(1+2\tilde\alpha e^{-2f}(g')^2\right) .
\end{equation}
We may split up the integrand as
\begin{equation}
    \left(\frac{-e^{f + (D-2) g}}{\left(f'+(D-2) g'\right)}\left(1+2\tilde\alpha e^{-2f}(g')^2\right)\right)\left(\frac{-\left(f'+(D-2) g'\right)\mathcal{F}^2}{\sqrt{1-\mathcal{F}^2}}\right)
\end{equation}
such that the right term is a total derivative. Conveniently, the left term can be identified as
\begin{equation}
    c_{\text{NEC}}(z) = \frac{-e^{f + (D-2) g}}{\left(f'+(D-2) g'\right)}\left(1+2\tilde\alpha e^{-2f}(g')^2\right)\; , \label{eqn:cnecfromEE}
\end{equation}
the NEC-motivated $c$-function \eqref{k=0 c func}. This $c$-function follows the coefficient constraints presented in \eqref{eq:cfnctCoeffs}, and further constrains $a_3=2$, where $a_3$ was previously free. However, \eqref{eqn:cnecfromEE} does not give $a_\text{UV}$ as its residue, since it does not follow \eqref{eqn:UVacons}. The expression for $c_\text{EE}$ can then be integrated by parts:
\begin{equation} \label{c_mono_arbitrary_D}
     c_{\text{EE}}(z_0) \propto -\sqrt{1-\mathcal{F}^2} \;c_{\text{NEC}}(z)\Bigg |_0^{z_0} + \int_0^{z_0}\dd z\sqrt{1-\mathcal{F}^2}\left(\dv{c_{\text{NEC}}}{z}\right) .
\end{equation}
After differentiating with respect to $z_0$, the surface term disappears since $\mathcal{F}(z_0)=1$. Similarly, the derivative hitting the upper integration bound gives no contribution. When computing $\dd c_{\text{EE}}/\dd z_0$ the $z_0$ derivative does not modify $\dd c_{\text{NEC}}/\dd z$, so the monotonicity of $c_{\text{NEC}}$ directly translates to monotonicity of $c_{\text{EE}}$.

\subsection{AdS$_{D+1}\to$ AdS$_{d+1}$ for general $d$}

One might also consider the more general case of flows down to AdS$_{d+1}$ with $d>2$. We will specialize our metric to be
\begin{equation}
    \dd s^2=e^{2f(z)}\qty(-\dd t+\dd z^2+\dd r^2+r^2 \dd\Omega_{d-2}^2)+e^{2g(z)}\dd s_{M_{D-d}}^2,
\end{equation}
and we will specify that our entangling region wraps $M_{D-d}$ and has a spherical cross-section of radius $R$. Given a profile $r(z)$, this then results in an induced metric
\begin{equation}
    \dd \sigma^2=e^{2f(z)}\qty(1+r'(z)^2)\dd z^2+e^{2f(z)}r(z)^2\dd\Omega_{d-2}^2+e^{2g(z)}\dd s_{M_{D-d}}^2,
\end{equation}
with induced Ricci scalar
\begin{align}
    \mathcal{R}=&(d-2)(d-3)\frac{e^{-2f}}{r^2}+(D-d)(D-d-1)\frac{\kappa}{\ell^2}e^{-2g}\nonumber\\
    &+\frac{e^{-2f}}{r^2(1+(r')^2)^2}\Big[-\qty(1+(r')^2)\Big(2(d-2)\qty((d-2)f'+(D-d)g')rr'+(d-2)(d-3)(r')^2\nonumber\\
    &+\big((d-2)(d-3)(f')^2+2(d-3)(D-d)f'g'+(D-d)(D-d-1)(g')^2+2(d-2)f''\nonumber\\
    &+2(D-d)g'')\big)r^2+2\qty(-(d-2)+\qty((d-2)f'+(D-d)g')rr')rr''\Big].
\end{align}
For more details see Appendix \ref{app:inducedR}. It is straightforward to check that for $d=2$, $\mathcal{R}$ reduces to \eqref{eq:Rind}. Similar to (\ref{eqn:AdS3JM}), the JM functional is then
\begin{align}
    S_\text{JM}=&\frac{\text{Vol}\qty(S^{d-2})\text{Vol}\qty(M_{D-d})}{4G_N}\int\dd{z}\Bigg\{r^{d-2} e^{(d-1)\tilde f}\sqrt{1+(r')^2}\qty(1+2\alpha(D-d)(D-d-1)\frac{\kappa}{\ell^2}e^{-2g})\nonumber\\
    &+2\alpha\, r^{d-4}\frac{e^{(d-1)\tilde f-2f}}{\sqrt{1+(r')^2}}\Big[r^2\big((d-2)(d-3)(f')^2+2(d-2)(D-d)f'g'\nonumber\\
    &+(D-d)(D-d-1)(g')^2\big)+2(d-2)rr'\qty((d-3)f'+(D-d)g')\nonumber\\
    &+(d-2)(d-3)\qty(1+2(r')^2)\Big]\Bigg\},\label{eqn:genJM}
\end{align}
where we have we have again integrated by parts and used the boundary term to cancel the Gibbons-Hawking term. If we set $\alpha=0$, this agrees with the two-derivative case \cite{GonzalezLezcano:2022mcd,Deddo:2022wxj}. Moreover, setting $D=d$, we recover
\begin{align}
    S_\text{JM}=&\frac{\text{Vol}\qty(S^{d-2})}{4G_N}\int\dd{z}r^{D-2} e^{(D-1)f}\sqrt{1+(r')^2}\left\{1+2\tilde\alpha\qty[\qty(f'+\frac{r'}{r})^2+\frac{1+(r')^2}{r^2}]\right\},\label{eqn:a=0JM}
\end{align}
which corresponds to the entanglement entropy of a spherical entangling region in flows within the same dimension, as studied in \cite{Hung:2011xb,deBoer:2011wk}.

However, the method applied in the $d=2$ case relied heavily on the fact that the integrand of $S_\text{JM}$ admitted a first integral. Since \eqref{eqn:genJM} contains an explicit factor of $r(z)$, one cannot use the same technique. Without a first integral, we cannot rewrite $r'(z)$ in terms of the turning point $z_0$ to produce an expression like (\ref{eq:EEcfnct}). Recall that monotonicity for the $d=2$ was demonstrated with respect to $z_0$, and it is not clear how one would proceed when $c_\text{EE}$ is not expressed as a function of $z_0$.


\section{Conclusions}\label{sec:end}

In this manuscript, we have explored higher-derivative renormalization group flows across dimensions. Our first look at holographic flows across dimensions involved explicitly constructing a $c$-function which is monotonically decreasing along flows from the UV to the IR as a consequence of the NEC. This $c$-function, just as the one constructed in the two-derivative case \cite{GonzalezLezcano:2022mcd}, is divergent; we have, however, shown that this divergence can be made to encode the UV central charge. Our second approach was to construct a monotonic $c$-function from the holographic entanglement entropy, which is given by a minimal surface prescription minimizing the Jacobson-Myers functional. We looked specifically at flows from AdS$_{D+1}$ to AdS$_3$ and explicitly constructed a monotonic $c$-function. More surprising is the fact that this $c$-function is related to the NEC-motivated  $c$-function.

Of course, one could ask: Given that the higher curvature corrections must be treated perturbatively, how could our story have failed? Considering that we are working perturbatively in $\alpha$, we can move terms from the numerator of our $c$-function \eqref{eq:LHcfnct} into the denominator, and so there are really only 5 free parameters to consider. On the other hand, the four-derivative part of NEC1 \eqref{eq:NEC1} has, up to our perturbative omission of terms proportional to $(g')^2$, 10 terms that must be matched in $c'(z)$. So, the fact that the NEC-motivated $c$-function evolves monotonically is a non-trivial statement  and could have easily not been the case.

We note that we could have additionally included in the action the quasi-topological term $\mathcal{Z}_{D+1}$ given by
\begin{align}
    \mathcal{Z}_{D+1}=&\hat R_{M\ \ N}^{\ \ \,P\ \ \, Q}\hat R_{P\ \,Q}^{\ \ R\ \ S}\hat R_{R\ \ S}^{\ \ M\ \ N}+\frac{1}{(2D-1)(D-3)}\bigg[\frac{3(3D-5)}{8}\hat R_{MNPQ}\hat R^{MNPQ}\hat R\nonumber\\
    &-3(D-1)\hat R_{MNPQ}\hat R^{MNP}_{\ \ \ \ \ \ R}\hat R^{QR}+3(D-1)\hat R_{MNPQ}\hat R^{MP}\hat R^{NQ}
    \nonumber\\
    &+6(D-1)\hat R_{MN}\hat R^{NP}\hat R_{P}^{\ \ \,M}-\frac{3(3D-1)}{2}\hat R_{MN}\hat R^{MN} \hat R+\frac{3(D-1)}{8}\hat R^3\bigg],
\end{align}
which was constructed in \cite{Myers:2010ru,Oliva:2010eb}; this term played a prominent role in \cite{Myers:2010xs,Myers:2010tj}. For our purposes, however, this term presents some difficulties. In contrast to the Gauss-Bonnet term, or even more generally Lovelock terms, the coefficients of $\mathcal{Z}_{D+1}$ are dimension-dependent. This presents us with a problem: We must either choose $\mathcal{Z}_{D+1}$, which is quasi-topological in the UV but which yields unsavory terms in the IR, or we could choose $\mathcal{Z}_{d+1}$ which is quasi-topological in the IR but not the UV.  The conundrum originates from the fact that $\mathcal{Z}_{D+1}$ was engineered to be quasi-topological for maximally symmetric backgrounds, and our background \eqref{Eq:Flowfg} does not satisfy this criterion. Hence, we would generically have to deal with fourth-order derivatives in the NEC.

As mentioned in the introduction and summary section, the field theory techniques required for proving monotonicity theorems are very dimension dependent. Recall that Zamolodchikov's proof of the $c$-theorem in 2d relied on properties of the correlator of two stress-energy tensors \cite{Zamolodchikov:1986gt} while in 4d Schwimmer and Komargodski relied on properties of certain four-point amplitudes to prove the $a$-theorem \cite{Komargodski:2011vj} . The entropic approach, due largely to Casini and collaborators, relied  almost exclusively  on strong subadditivity of the relative entropy \cite{Casini:2017vbe}. It is an outstanding problem to connect these different approaches. Holography has furnished two sets of proofs, one following from NEC and another related to the entropy via the Ryu-Takayanagi prescription. We have found, at least in a particular case, that the proofs are connected. We hope to explore this connection in more detail in the future and hope to draw lessons that might translate to field theoretic approaches. Another question that seems particularly suitable for  holographic attacks is the nature of supersymmetric flows; in this case, Einstein's equations can be replaced by a set of linear differential equations.

\section*{Acknowledgements}
We are grateful to Alfredo Gonz\'alez Lezcano for comments. This work is partially supported by the U.S. Department of Energy under grant DE-SC0007859. LPZ acknowledges support from an IBM Einstein Fellowship at the Institute for Advanced Study. ED and RJS were supported in part by Leinweber Graduate Summer Fellowships.

\appendix 
\section{Technical details}\label{App:Geom}
In this Appendix, we provide some supplementary technical details that were omitted from the main text.

\subsection{Riemann tensors}
\label{app:Riemann}
Here we collect Riemann tensors for the metric
\begin{equation}
    \dd s^2 = e^{2f(z)}(\eta_{\mu\nu}\dd x^\mu \dd x^\nu+ \dd z^2)+ e^{2g(z)}g_{ij}(y)\dd y^i\dd y^j.
\end{equation}
We will use $\mu,\nu,\rho,...$ for curved indices in the $d$-simensional base space and $i,j,k,...$ for curved indices on $M_{D-d}$, as well as $\alpha,\beta,\gamma,...$ for rigid indices in the $d$-dimensional spacetime and $a,b,c,d,...$ for rigid indices in the compact directions. We will use $z$ to denote the curved $z$-direction index and $\ul z$ to denote the rigid $z$-direction index. We will use $M, N,...$ for curved indices and $A,B,C,...$ for rigid indices of the whole $(D+1)$-dimensional spacetime. We choose a vielbein
\begin{align}
    \hat{e}^\alpha=e^{f(z)} e^\alpha,\ \ \ \hat e^{\ul z}=e^{f(z)}\dd z,\ \ \ \hat{e}^a=e^{g(z)} \tilde e^a,
\end{align}
so that $\dd s^2=\eta_{\alpha\beta}\hat e^\alpha\hat e^\beta+\hat e^{\ul z}\hat e^{\ul z}+\delta_{ab}\hat e^a\hat e^b$. Here we have defined $e^\alpha$ to be a vielbein for the flat $d$-dimensional space with metric $\eta_{\mu\nu}$ and $\tilde e^a$ to be a vielbein on $M_{D-d}$. Imposing the torsion-free condition
\begin{equation}
    \dd\hat e^A+\hat{\omega}^{A}_{\ \, B}\hat e^B=0,
\end{equation}
gives a spin connection
\begin{subequations}
    \begin{align}
    \hat\omega^{\alpha\beta}&=\omega^{\alpha\beta},\\
    \hat\omega^{\alpha\ul z}&=e^{-f}\partial^{\ul z}f\hat e^{\alpha},\\
    \hat\omega^{a \beta}&=0,\\
    \hat\omega^{a \ul z}&=e^{-f}\partial_{\ul z} g\,\hat e^a,\\
    \hat\omega^{ab}&=\tilde\omega^{ab},
    \end{align}
\end{subequations}
where $\omega$ is the spin connection on the $d$-dimensional base space and $\tilde\omega$ is the spin connection on $M_{D-d}$. The Riemann curvature two-form is then given by
\begin{equation}
    \hat R^{AB}=\dd\hat\omega^{AB}+\hat\omega^A_{\ \, C}\land\hat\omega^{CB},
\end{equation}
which, in components, reads
\begin{subequations}
\begin{align}
    \hat R^{\alpha\beta}_{\ \ \ \gamma\delta}&=-2e^{-2f}(f')^2\delta^{\alpha}_{[\gamma}\delta^{\beta}_{\delta]},\\
    \hat R^{\alpha\beta}_{\ \ \ \gamma\ul z}&=0,\\
    \hat R^{\alpha\ul z}_{\ \ \ \gamma\ul z}&=-e^{-2f}f''\delta^{\alpha}_{\gamma},\\
    \hat R^{\alpha b}_{\ \ \ \gamma d}&=-e^{-2f}f'g'\delta^{\alpha}_{\gamma}\delta^b_d,\\
    \hat R^{\alpha b}_{\ \ \ c \ul z}&=0,\\
    \hat R^{a\ul z}_{\ \ \ c\ul z}&=-e^{-2f}\qty(g''-f'g'+(g')^2)\delta^a_c,\\
    \hat R^{a b}_{\ \ \ c d}&=e^{-2g}\tilde R^{a b}_{\ \ \ c d}-2e^{-2f}(g')^2\delta^a_{[c}\delta^b_{d]},
\end{align}
\end{subequations}
where we have denoted the Riemann tensor on $M_{D-d}$ by $\tilde R^{a b}_{\ \ \ c d}$. Note that in the above, we have used the fact that the $d$-dimensional base space is flat to remove all the corresponding curvature tensors, hence why there is no $R_{\alpha\beta\gamma\delta}$. From here, one can compute the Ricci tensor, $\hat R_{AB}=\hat R^C_{\ \, ACB}$, to be
\begin{subequations}
    \begin{align}
        \hat R_{\alpha\beta}&= -e^{-2f}\qty[f''+(d-1)(f')^2+(D-d)f'g']\eta_{\alpha\beta},\\
        \hat R_{\alpha\ul z}&=0,\\
        \hat R_{\ul z\ul z}&=-e^{-2f}\qty[df''+(D-d)\qty(g''-g'f'+(g')^2)],\\
        \hat R_{a\beta}&=0,\\
        \hat R_{a\ul z}&=0,\\
        \hat R_{ab}&=e^{-2g}\tilde R_{ab}-e^{-2f}\qty[g''+(D-d)(g')^2+(d-1)f'g']\delta_{ab},
    \end{align}
\end{subequations}
where $\tilde R_{ab}$ denotes the Ricci tensor on $M_{D-d}$. Finally, the Ricci scalar is given by
\begin{align}
    \hat R=&e^{-2g}\tilde R-e^{-2f}\Big[2d f''+2(D-d)g''+d(d-1)(f')^2+2(d-1)(D-d)f'g'\nonumber\\
    &+(D-d+1)(D-d)(g')^2\Big],
\end{align}
where $\tilde R$ denotes the Ricci scalar on $M_{D-d}$.

\subsection{The general expression for $c'(z)$}
\label{app:c'(z)}

In section~\ref{sec:GBad}, we made an ansatz for a candidate $c$-function, (\ref{eq:LHcfnct}), in terms of real parameters $\{a_i,b_j\}$.  Given this ansatz, we find
\begin{align}
    c'(z)=\fft{e^{-\tilde f}(L_{\mathrm{eff}})^d}{G_N}\biggl\{&-(d-1)(f''-(f')^2)-(D-d)(g''+g'(g'-2f'))+\fft{(D-1)(D-d)}{d-1}(g')^2\nonumber\\
    &+\alpha e^{-2f}\Bigl[f''\left(\xi_1(f')^2+\xi_2f'g'+\xi_3 (g')^2\right)+g''\left(\xi_4(f')^2+\xi_5f'g'+\xi_6(g')^2\right)\nonumber\\
    &\kern1.5cm+\xi_7(f')^4+\xi_8(f')^3g'+\xi_9(f')^2(g')^2+\xi_{10}f'(g')^3+\xi_{11}(g')^4\Bigr]\nonumber\\
    &+\alpha e^{-2g}\frac{\kappa}{\ell^2}\Bigl[\omega_1f''+\omega_2g''+\omega_3(f')^2+\omega_4f'g'+\omega_5(g')^2\Bigr]\biggr\},
\label{eq:cprime}
\end{align}
where, for brevity, we have defined coefficients
{\allowdisplaybreaks
\begin{align}
    \xi_1&=-(d-3)a_1+3(d-1)a_4,\nn\\
    \xi_2&=2\frac{D-d}{d-1}a_1-(d-2)a_2+2(d-1)a_5,\nn\\
    \xi_3&=(d-1)(a_6-a_3)+\frac{D-d}{d-1}a_2,\nn\\
    \xi_4&=-(D-d)a_1+a_2+(d-1)a_5,\nn\\
    \xi_5&=-\frac{(D-d)(d-2)}{d-1}a_2+2a_3+2(d-1)a_6,\nn\\
    \xi_6&=-\frac{(D-d)(d-3)}{d-1}a_3+3(d-1)a_7,\nn\\
    \xi_7&=(d-3)a_1-3(d-1)a_4,\nn\\
    \xi_8&=2\frac{(D-d)(d-2)}{d-1}a_1+(d-3)a_2-(D-d)a_4-3(d-1)a_5,\nn\\
    \xi_9&=\frac{(D-d)\qty((D-d)a_1+2(d-2)a_2)}{d-1}+(d-3)a_3-(D-d)a_5-3(d-1)a_6,\nn\\
    \xi_{10}&=\frac{(D-d)\qty((D-d)a_2+2(d-2)a_3)}{d-1}-(D-d)a_6-3(d-1)a_7,\nn\\
    \xi_{11}&=(D-d)\qty(\frac{D-d}{d-1}a_3-a_7),\nn\\
    \omega_1&=-(d-1)(b_1-b_2),\nn\\
    \omega_2&=(d-1)b_3-(D-d)b_1,\nn\\
    \omega_3&=(d-1)(b_1-b_2),\nn\\
    \omega_4&=2(D-d-1)b_1-(D+d-2)b_2-(d-1)b_3,\nn\\
    \omega_5&=\frac{(D-d)(D-d-2)}{d-1}b_1-(D+d-2)b_3.
\end{align}
}%
Note that the form of the ansatz, (\ref{eq:LHcfnct}), was chosen so that no higher than second derivatives of $f$ and $g$ appear in (\ref{eq:cprime}).


\subsection{Induced Ricci scalar}\label{app:inducedR}
In Section \ref{sec:EE}, we require an expression for Ricci scalar of the induced metric on the entangling surface, which we compute here. The induced metric is given by
\begin{equation}
    \dd \sigma^2=e^{2f(z)}\qty(1+r'(z)^2)\dd z^2+e^{2f(z)}r(z)^2\dd\Omega_{d-2}^2+e^{2g(z)}\dd s_{M_{D-d}}^2.
\end{equation}
By slight abuse of notation, we will use $\alpha,\beta,\gamma,\delta,...$ to index the rigid indices along the unit $(d-2)$-sphere (for this section only, these indices will not run over $t$ or $r$). A natural choice of vielbein is then
\begin{equation}
    \hat e^{\bar z}=e^f\sqrt{1+(r')^2}\dd z,\ \ \hat e^\alpha=e^fre^\alpha,\ \ \hat e^a=e^g\tilde e^a,
\end{equation}
where $e^\alpha$ is a vielbein on the $(d-2)$-sphere and $\tilde e^a$ is a vielbein on $M_{D-d}$. Note that this notation differs from the previous subsection. As before, we make use of the torsion-free condition to compute the components of the spin connection
\begin{subequations}
    \begin{align}
        \hat\omega^{\alpha\beta}&=\omega^{\alpha\beta},\\
        \hat\omega^{\alpha\ul z}&=\frac{e^{-f}}{\sqrt{1+(r')^2}}\qty(f'+\frac{r'}{r})\hat e^\alpha,\\
        \hat\omega^{\alpha b}&=0,\\
        \hat\omega^{ab}&=\tilde\omega^{ab},\\
        \hat\omega^{a\ul z}&=\frac{g'\,e^{-f}}{\sqrt{1+(r')^2}},
    \end{align}
\end{subequations}
where $\omega$ is the spin connection on the $(d-2)$-sphere and $\tilde\omega$ is the spin connection on $M_{D-d}$. The  induced Riemann tensor components may then be computed to be
\begin{subequations}
    \begin{align}
        \cal{R}^{\alpha\beta}_{\ \ \ \gamma\delta}&=\frac{e^{-2f}}{r^2}\bar R^{\alpha\beta}_{\ \ \ \gamma\delta}-2\frac{e^{-2f}}{1+(r')^2}\qty(f'+\frac{r'}{r})^2\delta^{[\alpha}_\gamma\delta^{\beta]}_\delta,\\
        \mathcal{R}^{\alpha\ul z}_{\ \ \ \beta\ul z}&=-\qty[\dv{}{z}\qty(\frac{e^{-f}}{\sqrt{1+(r')^2}}\qty(f'+\frac{r'}{r}))\frac{e^{-f}}{\sqrt{1+(r')^2}}+\frac{e^{-2f}}{1+(r')^2}\qty(f'+\frac{r'}{r})^2]\delta^\alpha_\beta,\\
        \mathcal{R}^{a\ul z}_{\ \ \ b\ul z}&=-\qty[\dv{}{z}\qty(\frac{g'\,e^{-f}}{\sqrt{1+(r')^2}})\frac{e^{-f}}{\sqrt{1+(r')^2}}+\frac{e^{-2f}(g')^2}{1+(r')^2}]\delta^a_b,\\
        \mathcal{R}^{ab}_{\ \ \ cd}&=e^{-2g}\tilde R^{ab}_{\ \ \ cd}-2\frac{e^{-2f}(g')^2}{1+(r')^2}\delta^{[a}_c\delta^{b]}_d,\\
        \mathcal{R}^{\alpha b}_{\ \ \ \gamma d}&=-\frac{g'\, e^{-2f}}{1+(r')^2}\qty(f'+\frac{r'}{r})\delta^a_c\delta^\beta_\delta,
    \end{align}\end{subequations}
    where $\bar R^{\alpha\beta}_{\ \ \ \gamma\delta}$ denotes the Riemann tensor on the (unit) $(d-1)$-sphere and $\tilde R^{ab}_{\ \ \ cd}$ denotes the Riemann tensor on $M_{D-d}$. Computing the induced Ricci scalar as $\mathcal{R}=\mathcal{R}^{AB}_{\ \ \ \,AB}$, and using the identities for the Ricci scalars of the constituent metrics
    \begin{subequations}
        \begin{align}
            \bar R&=(d-2)(d-3),\\
            \tilde R&=(D-d)(D-d-1)\frac{\kappa}{\ell^2},
        \end{align}
    \end{subequations}
    we finally arrive at our expression for the induced Ricci scalar
    \begin{align}
    \mathcal{R}=&(d-2)(d-3)\frac{e^{-2f}}{r^2}+(D-d)(D-d-1)\frac{\kappa}{\ell^2}e^{-2g}\nonumber\\
    &+\frac{e^{-2f}}{r^2(1+(r')^2)^2}\Big[-\qty(1+(r')^2)\Big(2(d-2)\qty((d-2)f'+(D-d)g')rr'+(d-2)(d-3)(r')^2\nonumber\\
    &+\big((d-2)(d-3)(f')^2+2(d-3)(D-d)f'g'+(D-d)(D-d-1)(g')^2+2(d-2)f''\nonumber\\
    &+2(D-d)g'')\big)r^2+2\qty(-(d-2)+\qty((d-2)f'+(D-d)g')rr')rr''\Big].
\end{align}

\section{Alternate coordinates}
Here we collect some of the results from the main text reexpressed in alternate coordinates, more akin to those used in \cite{Myers:2010xs,Myers:2010tj,Myers:2012ed}. These are not new results, but the reader might find them more useful for some purposes.

\subsection{NEC-motivated $c$-function}\label{app:altCoords1}
One may alternately parameterize the metric as
\begin{equation}
    \dd{s}^2=e^{2A(r)}\eta_{\mu\nu}\dd{x}^\mu \dd{x}^\nu+\dd{r}^2+e^{2B(r)}g_{ij}(y)\dd{y}^i\dd{y}^j.\label{eq:altCoords}
\end{equation}
These are the coordinates that are used in \cite{Myers:2010tj,Myers:2010xs}. Pure AdS corresponds to $A(r)=B(r)=r/L$, and so it is natural to identify $r=0$ with the IR and $r=\infty$ with the UV. We expect the asymptotic behavior of the metric functions to be
\begin{align}
    &r\to\infty:\ &A(r)\to \frac{r}{L_\text{UV}},&\qquad B(r)\to \frac{r}{L_\text{UV}},\nonumber\\
    &r\to0:\ &A(r)\to \frac{r}{L_\text{IR}},&\qquad B(r)\to B_\text{IR}.
\end{align}
We still assume that the internal manifold is maximally symmetric with Ricci scalar
\begin{equation}
    \tilde R=(D-d)(D-d-1)\frac{\kappa}{\ell^2}.
\end{equation}
One can take this metric and compute the resulting $t$-$z$ null energy condition NEC1 for arbitrary dimensions, which gives
\begin{align}\label{NEC1 arbitrary}
    0\le& -(d-1)A''-(D-d)B''+(D-d)A'B'-(D-d)(B')^2\nonumber\\
    &+\alpha\Big[2(d-1)(d-2)(d-3)(A')^2A''+4(d-1)(d-2)(D-d)A'B'A''\nonumber\\
    &+2(d-1)(D-d)(D-d-1)(B')^2A''+2(d-1)(d-2)(D-d)(A')^2B''\nonumber\\
    &+4(d-1)(D-d)(D-d-1)A'B'B''+2(D-d)(D-d-1)(D-d-2)(B')^2B''\nonumber\\
    &-2(d-1)(d-2)(D-d)(A')^3B'-2(d-1)(D-d)(2D-3d)(A')^2(B')^2\nonumber\\
    &-2(D-d)(D-d-1)(D-3d)A'(B')^3+2(D-d)(D-d-1)(D-d-2)(B')^4\Big]\nonumber\\
    &+2\alpha(D-d)(D-d-1)\frac{\kappa}{\ell^2}\qty[-(d-1)A''+(D-d-2)\qty(-B''+A'B'-(B')^2)].
\end{align}
One might then propose a generic candidate $c$-function
\begin{align}
    c(r)=\frac{e^{(D-d)B}\qty(1+\alpha\qty(a_1 (A')^2+a_2A'B'+a_3(B')^2+b_1\frac{\kappa}{\ell^2}e^{-2B}))}{\qty(\tilde{A}'+\alpha\qty(a_4(A')^3+a_5(A')^2B'+a_6A'(B')^2+a_7(B')^3+b_2\frac{\kappa}{\ell^2}e^{-2B}A'+b_3\frac{\kappa}{\ell^2}e^{-2B}B'))^{d-1}},
\end{align}
where we have defined
\begin{equation}
    \tilde{A}=A+\frac{D-d}{d-1}B,
\end{equation}
in analogy to $\tilde f$. This $c$-function is the obvious generalization of the two-derivative case (when $\alpha=0$). As before, one computes
\begin{align}
    c'(z)=&\frac{e^{(D-d)B}}{\qty(\tilde{A}'+\alpha\qty(a_3(A')^2B'+a_4A'(B')^2+a_5(B')^3+b_2\frac{\kappa}{\ell^2}e^{-2B}A'+b_3\frac{\kappa}{\ell^2}e^{-2B}B'))^{d}}\nonumber\\
    &\times\Big\{-(d-1)A''-(D-d)B''+(D-d)A'B'+(D-d)^2(B')^2\nonumber\\
    &+\alpha\Big[\xi_1(A')^2A''+\xi_2A'B'A''+\xi_3(B')^2A''+\xi_4(A')^2B''+\xi_5A'B'B''+\xi_6(B')^2B''\nonumber\\
    &+\xi_7(A')^4+(A')^3B'+\xi_8(A')^2(B')^2+\xi_9A'(B')^3+\xi_{10}(B')^4\Big]\nonumber\\
    &+\alpha\frac{\kappa}{\ell^2}\qty[\omega_1A''+\omega_2B''+\omega_3A'B'+\omega_4(B')^2]\Big\},
\end{align}
where we have defined
\begin{align}
    \xi_1&=3(a_1+a_4)-(a_1+3a_4)d,\nonumber\\
    \xi_2&=a_2-(d-1)(a_2+2a_5)+2\frac{D-d}{d-1}a_1,\nonumber\\
    \xi_3&=-(d-1)a_6+\frac{D-d}{d-1}a_2,\nonumber\\
    \xi_4&=a_2-(d-1)a_5-(D-d)a_1,\nonumber\\
    \xi_5&=-2(d-1)a_6+\frac{(D-d)(d-2)}{d-1}a_2,\nonumber\\
    \xi_6&=-3(d-1)a_7,\nonumber\\
    \xi_7&=(D-d)(a_1+a_4),\nonumber\\
    \xi_8&=(D-d)\qty(a_2+a_5+\frac{D-d}{d-1}a_1),\nonumber\\
    \xi_9&=(D-d)\qty(a_6+\frac{D-d}{d-1}a_2),\nonumber\\
    \xi_{10}&=(D-d)a_7,\nonumber\\
    \omega_1&=-(d-1)b_2,\nonumber\\
    \omega_2&=-(d-1)b_3,\nonumber\\
    \omega_3&=(D-d-2)b_1+(D-d-2)b_2,\nonumber\\
    \omega_4&=(D+d-2)b_3+\frac{(D-d)(D-d-2)}{d-1}b_1.
\end{align}
With the particular choice of
\begin{subequations}
    \begin{align}
        a_1&=-2(d-1)(d-2),\\
        a_2&=-4(D-d)(d-2),\\
        a_4&=0,\\
        a_5&=-4\frac{(D-d)(d-2)}{(d-1)},\\
        a_6&=\frac{a_2}{d-1}+2\frac{D-d}{(d-1)^2}(1+d(-5+3d-2D)+3D),\\
        b_1&=\frac{2(D-d-1)((D+1)d-D-d^2+2)}{d},\\
        b_2&=\frac{2(D-d-1)(D-d-2)}{d},\\
        b_3&=\frac{2(D-d)(D-d-1)(D-3d-2)}{d(d-1)},
    \end{align}
\end{subequations}
we get that
\begin{equation}
    c'(r)=\frac{e^{(D-d)B}\qty(\text{NEC1}+\frac{(D-1)(D-d)}{d-1}(B')^2\qty(1+\mathcal{O}(\alpha)))}{\qty(\tilde{A}'+\alpha\qty(a_3(A')^2B'+a_4A'(B')^2+a_5(B')^3+b_2\frac{\kappa}{\ell^2}e^{-2B}A'+b_3\frac{\kappa}{\ell^2}e^{-2B}B'))^{d}}\ge 0,
\end{equation}
and hence the candidate $c$-function gives us a monotonic flow from the UV to the IR. 

As before, we never need to use the all-internal components of the Riemann tensor $\hat R_{ijkl}$ to obtain NEC1, and so the above results also trivially generalize to arbitrary Einstein internal manifolds, as in the $f$ and $g$ coordinates. 

\subsection{Entanglement entropy $c$-function}\label{app:altCoords2}
One may also repeat the arguments of Section \ref{sec:EEcfnct} in the alternate coordinates \eqref{eq:altCoords}. Here we focus on flows from AdS$_{D+1}$ to AdS$_3$, and so we specialize the metric \eqref{eq:altCoords} to
\begin{equation}
    \dd \sigma^2=e^{2A(r)}\qty(-\dd t^2+\dd \rho^2)+\dd r^2+e^{2B(r)}\dd s_{M_{D-2}}^2.
\end{equation}
In terms of a profile $\rho(r)$, the induced Ricci scalar is
\begin{align}
    \mathcal{R}=&-\frac{(D-2)}{\left(1+e^{2 A(r)} \rho '(r)^2\right)^2}\Big(e^{2 A(r)} \rho '(r) \left(\rho '(r) \left(2 A'(r) B'(r)-(D-1) B'(r)^2-2 B''(r)\right)+2 B'(r) \rho ''(r)\right)\nonumber\\
    &-(D-1) B'(r)^2-2 B''(r)\Big)+(D-2)(D-3)\frac{\kappa}{\ell^2}e^{-2B},
\end{align}
which leads to a JM functional whose first integral is
\begin{equation}
    C=\frac{\rho'(r) e^{2 A(r)+(D-2)B(r)} \left(\left(1+e^{2 A(r)} \rho '(r)^2\right) \left(1+2 \tilde\alpha \frac{\kappa}{\ell^2}e^{-2B} \right)-2\tilde\alpha B'(r)^2\right)}{\left(1+e^{2 A(r)} \rho '(r)^2\right)^{3/2}},
\end{equation}
which can be solved to give
\begin{equation}
    \rho'(r)=\frac{e^{-A}\mathcal{F}}{\sqrt{1-\mathcal{F}^2+4\tilde\alpha\qty(\frac{\kappa}{\ell^2}e^{-2B}-(B')^2(1-\mathcal{F}^2))}},\ \ \mathcal{F}(r)\equiv C e^{-A-(D-2)B}.
\end{equation}
To fix the value of $C$, we note that we should have $\rho'(r)\to-\infty$ as $r\to r_0$, where $r_0$ is the deepest point in the bulk that the minimal surface. This then requires that
\begin{equation}
    C=e^{A(r_0)+(D-2)B(r_0)}\qty(1+2\tilde\alpha\frac{\kappa}{\ell^2}e^{-2B(r_0)}).
\end{equation}
Then the radius of the entangling area is
\begin{align}
    R=&\int_{r_0}^\infty\dd{r}\rho'(r)=\int_{r_0}^\infty \dd{r}\frac{e^{-A}\mathcal{F}}{\sqrt{1-\mathcal{F}^2+4\tilde\alpha\qty(\frac{\kappa}{\ell^2}e^{-2B}-(B')^2(1-\mathcal{F}^2))}}\nonumber\\
    =&\int_{r_0}^\infty \dd{r}\qty[\frac{e^{-A}\mathcal{F}}{\sqrt{1-\mathcal{F}^2+4\tilde\alpha\frac{\kappa}{\ell^2}e^{-2B}}}+2\tilde\alpha \frac{e^{-A}(B')^2\mathcal{F}}{\sqrt{1-\mathcal{F}^2}}]+\mathcal{O}(\tilde\alpha^2)\nonumber\\
    =&\lim_{r_c\to\infty}\int_{r_0}^{r_c}\dd{r}\Big[\sqrt{1-\mathcal{F}^2+4\tilde\alpha\frac{\kappa}{\ell^2}e^{-2B}}\dv{}{r}\frac{e^{-A}}{\mathcal{F}'+\frac{4\tilde\alpha}{\mathcal{F}}\frac{\kappa}{\ell^2}e^{-2B}B'}\nonumber\\
    &+2\tilde\alpha \sqrt{1-\mathcal{F}^2}\dv{}{r}\qty(\frac{e^{-A}(B')^2}{\mathcal{F}'})\Big]+2\tilde\alpha\lim_{r_c\to\infty}\frac{e^{-A}(B')^2}{\mathcal{F}'}\Bigg\vert_{r=r_c}+\mathcal{O}(\tilde\alpha^2).
\end{align}
Equivalently, in terms of a profile $r(\rho)$, we may write
\begin{equation}
    r'(\rho)=e^{A}\frac{\sqrt{1-\mathcal{F}^2+4\tilde\alpha\qty(\frac{\kappa}{\ell^2}e^{-2B}-(B')^2(1-\mathcal{F}^2))}}{\mathcal{F}}.
\end{equation}
The JM functional may then be calculated as
\begin{align}
    S_\text{JM}=&\frac{2\text{Vol}(M_{D-2})}{4G_N}\int\dd{\rho}\qty[\sqrt{e^{2A}+(r')^2}e^{(D-2)B}\qty(1+2\tilde\alpha\frac{\kappa}{\ell^2}e^{-2B})+2\tilde\alpha\frac{e^{(D-2)B}(B')^2(r')^2}{\sqrt{e^{2A}+(r')^2}}]\nonumber\\
    &-4\tilde\alpha e^{(D-2)B}B'\Big\vert_{\rho=\rho_c}.
\end{align}
As before, the boundary term is independent of $R$ and so it will not affect the succeeding analysis. The monotonic central charge is then given by
\begin{equation}
    c_\text{EE}=R\partial_R S_\text{JM}=\frac{2\pi\text{Vol}(M_{D-2})}{\ell_P^{D-1}}e^{A_0+(D-2)B_0}\qty(1+2\tilde\alpha\frac{\kappa}{\ell^2}e^{-2B_0})R.
\end{equation}

As before, this generically leads to rather complicated terms
\begin{align*}
    \dv{c_\text{EE}}{r_0}=&\frac{2\text{Vol}\qty(M_{D-2})}{4G_N}\int\dd\rho\frac{e^{(D-2)B}\mathcal{F}^2\tilde A_0'}{\sqrt{1-\mathcal{F}^2}(\tilde A')^2}\Bigg\{-A''-(D-2)B''+(D-2)A'B'+(D-2)^2(B')^2\nonumber\\
    &+2\tilde\alpha B'\qty[A'((D-2)(B')^2+2B'')+B'(-A''+(D-2)((D-2)(B')^2+B'')]\nonumber\\
    &+\tilde\alpha\frac{\kappa}{\ell^2}\Bigg[e^{\tilde A-2B}\frac{A''+(D-2)(B''-\tilde A' B')}{e^{\tilde A}-e^{\tilde A_0}}+4e^{-2B_0}B_0'\frac{A''+(D-2)(B''-\tilde A' B')}{\tilde A_0'}\nonumber\\
    &-\frac{4e^{2(\tilde A-2\tilde A_0-2B-B_0)}\qty(e^{2\tilde A_0}-2e^{\tilde A})}{\tilde A'}\nonumber\\
    &\times\qty(A'(2\tilde A'B'+B'')+B'((3D-8)\tilde A'B'-2A''-(D-2)B''))\Bigg]\Bigg\},
\end{align*}
but, if one sets $\kappa=0$, then
\begin{equation}
    \dv{c_\text{EE}}{r_0}=\frac{2\text{Vol}\qty(M_{D-2})}{4G_N}\int\dd\rho\frac{e^{(D-2)B}\mathcal{F}^2\tilde A_0'}{\sqrt{1-\mathcal{F}^2}(\tilde A')^2}\qty[\text{NEC1}+(D-1)(D-2)(B')^2\qty(1+\mathcal{O}(\tilde\alpha))]\ge0,
\end{equation}
which gives monotonicity along flows to the IR. This parallels the computation that was done in $f$ and $g$ coordinates.

\bibliographystyle{JHEP}
\bibliography{IrreversibilityRefs.bib}

\end{document}